# A Neural Network-Based Enrichment of Reproducing Kernel Approximation for Modeling Brittle Fracture

Jonghyuk Baek[1] and J. S. Chen[1*]

[1]Department of Structural Engineering, University of California, San Diego

## Abstract

Numerical modeling of localizations is a challenging task due to the evolving rough solution in which the localization paths are not predefined. Despite decades of efforts, there is a need for innovative discretization-independent computational methods to predict the evolution of localizations. In this work, an improved version of the neural network-enhanced Reproducing Kernel Particle Method (NN-RKPM) is proposed for modeling brittle fracture. In the proposed method, a background reproducing kernel (RK) approximation defined on a coarse and uniform discretization is enriched by a neural network (NN) approximation under a Partition of Unity framework. In the NN approximation, the deep neural network automatically locates and inserts regularized discontinuities in the function space. The NN-based enrichment functions are then patched together with RK approximation functions using RK as a Partition of Unity patching function. The optimum NN parameters defining the location, orientation, and displacement distribution across location together with RK approximation coefficients are obtained via the energy-based loss function minimization. To regularize the NN-RK approximation, a constraint on the spatial gradient of the parametric coordinates is imposed in the loss function. Analysis of the convergence properties shows that the solution convergence of the proposed method is guaranteed. The effectiveness of the proposed method is demonstrated by a series of numerical examples involving damage propagation and branching.

## 1. Introduction

Neural networks (NNs) have been shown to have powerful approximation ability [1,2]. The strong adaptivity and hidden information extraction capability have made deep neural networks a core element of machine learning in various applications. This feature also makes NNs appealing for solving challenging problems in computational mechanics. For example, data-driven computations for path-dependent material modeling [3–8], reduced order modeling [9,10], and parameter identification [11–13]. Additionally, the flexible adaptivity in NN allows an approximation space to be goal-specifically optimized. Utilizing this flexibility in the approximation space, NNs can be considered an alternative to traditional mesh-based methods in solving challenging problems involving localizations, such as fracture, for which special treatment is needed near the localizations.

---

[*] Corresponding author.
E-mail address: jsc137@ucsd.edu

Traditional approaches for fracture modeling can be divided into two broad categories: discrete crack approaches and diffuse crack approaches. The former category includes extended or generalized FEMs [14–16], partition of unity-based enrichment [17,18], and meshfree method with near-tip enrichment [19,20]. In these methods, strong discontinuities are directly inserted into the approximation, necessitating the detection and tracking of crack surfaces, significantly increasing the complexity of the computation for multidimensional problems. Nonlocal averaging [21], high order gradient models [22–24], and phase field methods [25–28] have been employed in the diffuse crack approaches. In this family of methods, nonlocal effects are typically introduced in the approximation or in the energy function, yielding diffused, regularized representation of cracks. This property enables traditional mesh-based or meshfree methods to approximate localizations without enrichment and the need for localization tracking. However, for sufficient accuracy, intense mesh refinement is required in the regions of localizations. For example, Geelen et al. (2019) [28] used an element size as small as one-tenth the width of the diffuse crack.

With their adaptive nature as an approximation, NNs provide a new paradigm in searching for solutions of mathematical models. Recently, NNs have been successfully applied as a solver of partial differential equations [11,12,29–33]. In physics-informed neural network (PINN)by Raissi et al. ) [11,12], the solution of a PDE is approximated by densely-connected deep neural networks with the residual-based loss function minimization. Haghighat and Juanes (2021) [34] developed the Python package SciANN for scientific computing using PINN and demonstrated its ability to capture strain and stress localization in a perfectly plastic material. More recently, PINNs have been extended to multi-physics problems [35,36]. However, one drawback of utilizing a deep neural network combined with a residual-based and collocated loss function is its computational cost, e.g., in [34], where 100 million unknown weights and biases were used. Samaniego et al. (2020) [29] demonstrated that potential-based loss functions produced superior results with significantly fewer unknowns than the residual-based loss function commonly used in PINN. Zhang et al. (2021) [30] proposed a deep neural network that reproduces standard approximations along with automatic refinement enabled by treating nodal positions as unknown network parameters, which, however, introduces sparsity into the neural network. Lu et al. (2021) [31], based on the universal approximation theorem [37], designed a new deep neural network architecture, in which the output of one deep neural network is multiplied by the output of another deep neural network, resulting effective approximations of nonlinear operators in partial differential equations.

Despite the growing interest in PINNs, there has been limited research on developing effective and computationally efficient NN-based approximation for modeling localizations. Baek et al. (2022) [33] proposed a neural network-enhanced reproducing kernel particle method (NN-RKPM) for modeling localizations. In this work, the approximation is constructed as the superposition of the NN approximation and the reproducing kernel (RK) approximation. For computational efficiency, NNs are limited to approximating localizations, while the RK approximation on a coarse and uniform discretization is employed to approximate the smooth solutions. In this approach, the NN approximation control parameters play the role in automatically capturing the location, orientation, and the localization profile at the localizations. These NN parameters are determined by the optimization of an energy-based loss function. In this work, we propose an improved version of NN-RKPM in which the NN approximation and the background RK approximation are patched together with Partition of Unity for ensured convergence. This approach is derived through an NN-based correction of standard RK shape functions. In the modified NN-RK approximation, the deep neural network automatically locates and inserts regularized discontinuities in the function space, and the NN enriched RK coefficient function provides varying magnitude of the discontinuity along the localization path. Additionally, convergence properties of the proposed method are analyzed.

The paper is organized as follows. In Section 2, the basic equations are provided, including the minimization problem for brittle fracture and the reproducing kernel particle method. In Section 3, a neural network-enriched Partition of Unity reproducing kernel approximation is proposed, along with convergence analysis and regularization technique. In Section 4, the implementation details including the neural network architecture and solution procedure are provided. This is followed by numerical examples in Section 5 and concluding remarks in Section 6.

## 2. Background

### 2.1. Minimization Problem for Fracture

For a domain $\Omega \in \mathbb{R}^d$ with the space dimension $d$ and its boundary $\partial\Omega = \partial\Omega_g \cup \partial\Omega_h$ that consists of the Dirichlet boundary $\partial\Omega_g$ and the Neumann boundary $\partial\Omega_h$, let us consider the following minimization problem: for $\mathbf{u} \in H^1$, $\mathbf{u} = \mathbf{g}$ on $\partial\Omega_g$,

$$\min_{\mathbf{u}} \Pi(\mathbf{u}) = \int_\Omega \psi(\mathbf{u}) \, d\Omega - \int_\Omega \mathbf{u} \cdot \mathbf{b} \, d\Omega - \int_{\partial\Omega_h} \mathbf{u} \cdot \mathbf{h} \, d\Gamma, \tag{1}$$

where $\mathbf{u}$, $\psi(\mathbf{u})$, $\mathbf{b}$, and $\mathbf{h}$ are the displacement, energy density functional, body force, and traction, respectively. The energy density functional $\psi(\mathbf{u})$ has the following form:

$$\psi(\mathbf{u}) = g\left(\eta(\boldsymbol{\varepsilon}(\mathbf{u}))\right)\psi_0^+(\mathbf{u}) + \psi_0^-(\mathbf{u}) + \bar{\psi}\left(\eta(\boldsymbol{\varepsilon}(\mathbf{u}))\right). \tag{2}$$

Herein, $\boldsymbol{\varepsilon} = \frac{1}{2}(\nabla \mathbf{u} + (\nabla \mathbf{u})^T)$, $\eta$, and $g$ are the strain tensor, the (strain dependent) damage variable, and the degradation function, respectively. Three energy density components $\psi_0^+$, $\psi_0^-$, and $\bar{\psi}$ denote non-degraded tensile strain energy, compressive strain energy, and dissipation functional, respectively. The tensile and compressive strain energies are defined as

$$\psi_0 = \mu \bar{\varepsilon}_i \bar{\varepsilon}_i + \frac{\lambda}{2} tr(\bar{\boldsymbol{\varepsilon}})^2,$$

$$\psi_0^+ = \mu \langle \bar{\varepsilon}_i \rangle_+ \langle \bar{\varepsilon}_i \rangle_+ + \frac{\lambda}{2} \langle tr(\bar{\boldsymbol{\varepsilon}}) \rangle_+^2, \tag{3}$$

$$\psi_0^- = \psi_0 - \psi_0^+,$$

where the summation notation is adopted. In (3), $\bar{\boldsymbol{\varepsilon}}$, $\lambda$, and $\mu$ are principal strain, Lamé's first and second parameters, respectively. $\langle \cdot \rangle_+ = \max(\cdot, 0)$ and $\langle \cdot \rangle_- = \min(\cdot, 0)$ are additionally used. The stress is defined as

$$\boldsymbol{\sigma} = g(\eta(\boldsymbol{\varepsilon})) \frac{\partial \psi_0^+}{\partial \boldsymbol{\varepsilon}} + \frac{\partial \psi_0^-}{\partial \boldsymbol{\varepsilon}}. \tag{4}$$

In this work, the damage variable, dissipation functional, and degradation function are defined as follows:

$$\eta = \frac{\psi_0^+}{\psi_0^+ + p} \tag{5}$$

$$\bar{\psi} = p\eta^2, \tag{6}$$

$$g = (1 - \eta)^2, \tag{7}$$

where $p$ is a fracture energy-dependent material property. The adopted dissipation functional and degradation function in Eqs. (6) and (7) are the same as what is used in Miehe et al. (2010)[25] except the absence of the higher order term $\mathcal{O}(\nabla \eta^2)$ in the dissipation functional in (6). Therefore, it is straightforward to show that the damage model in Eqs. (5)-(7) is variationally consistent, i.e., for $\mathbf{u} \in H^1$, $\mathbf{u} = \mathbf{g}$ on $\partial \Omega_g$, for all $\delta \mathbf{u} \in H^1$, $\delta \mathbf{u} = \mathbf{0}$ on $\partial \Omega_g$,

$$\delta \Pi = \int_\Omega \delta \boldsymbol{\varepsilon}(\mathbf{u}) : \boldsymbol{\sigma}(\boldsymbol{\varepsilon}) \, d\Omega = \int_\Omega \delta \mathbf{u} \cdot \mathbf{b} \, d\Omega + \int_{\partial \Omega^h} \delta \mathbf{u} \cdot \mathbf{h} \, d\Gamma, \tag{8}$$

which leads to the following balance equation:

$$\nabla \cdot \boldsymbol{\sigma} + \mathbf{b} = \mathbf{0} \text{ in } \Omega, \tag{9}$$

with the boundary conditions

$$\mathbf{u} = \mathbf{g} \text{ on } \partial \Omega_g, \tag{10}$$

$$\nabla \mathbf{u} \cdot \mathbf{n} = \mathbf{h} \text{ on } \partial \Omega_h, \tag{11}$$

where $\mathbf{n}$ denotes the surface normal vector.

To achieve the irreversibility of the damage, a history variable

$$\mathcal{H} = \max \left( \max_{t \in [0,T]} \{\psi_0^+(\boldsymbol{\varepsilon}) - \psi_c\}, 0 \right) \tag{12}$$

is employed to describe the damage variable:

$$\eta = \frac{\mathcal{H}}{\mathcal{H} + p}. \tag{13}$$

For Eq. (12), the critical fracture energy $\psi_c$ is defined as

$$\psi_c = \frac{f_t}{2E} \tag{14}$$

with the tensile strength of material $f_t$ and Young's modulus $E$. The model parameter $p$ takes the following form

$$p = \frac{\mathcal{G}_c}{\ell}, \tag{15}$$

with critical energy release rate $\mathcal{G}_c$ and length scale parameter $\ell$. To take mixed mode fracture into account, we adopt the $\mathcal{F}$-criterion[38], with the mode I critical energy release rate $\mathcal{G}_{cI}$ and the mode II critical energy release rate $\mathcal{G}_{cII}$:

$$\mathcal{F} \equiv \frac{\psi_0^+}{\mathcal{G}_c} \approx \frac{\psi_I^+}{\mathcal{G}_{cI}} + \frac{\psi_{II}^+}{\mathcal{G}_{cII}}, \tag{16}$$

with

$$\psi_I^+ = \frac{\lambda}{2} \langle \sum \bar{\varepsilon}_i \rangle_+^2, \tag{17}$$

$$\psi_{II}^+ = \mu \langle \bar{\varepsilon}_i \rangle_+ \langle \bar{\varepsilon}_i \rangle_+. \tag{18}$$

Eq. (16) leads to the following critical energy release rate:

$$\mathcal{G}_c = \frac{\psi_0^+}{\psi_I^+/\mathcal{G}_{cI} + \psi_{II}^+/\mathcal{G}_{cII}}. \tag{19}$$

Note that Eq. (19) implies $\mathcal{G}_c = \mathcal{G}_{cI}$ for pure mode I fracture when $\psi_0^+ = \psi_I^+$ and $\mathcal{G}_c = \mathcal{G}_{cII}$ for pure mode II fracture when $\psi_0^+ = \psi_{II}^+$.

***Remark 1.1.*** With $\mathcal{G}_c$ defined in (19) which is a function of strain, the functional $\Pi$ defined in (1) is not a minimization functional for the Euler-Lagrange equation (9). Therefore, in this work, we solve the minimization problem in (1) and the $\mathcal{G}_c$ calculation in (19) in a staggered manner.

***Remark 1.2.*** Different from the phase field fracture methods, the damage model described in this section is a local model in the absence of the higher order term in the dissipation functional. Therefore, there is possibility of the loss of ellipticity and the discretization-dependence of the numerical solution. This issue will be addressed in Section 3.3.

## 2.2. Reproducing kernel particle method for background approximation

Here we review the standard reproducing kernel particle method (RKPM) that is used to approximate smooth part of the solution in the proposed approach (see Section 3).

### 2.2.1. Reproducing kernel approximation

Let $\Omega$ be a domain discretized by $NP$ nodes with nodal coordinate $\{\mathbf{x}_I\}_{I \in \mathcal{S}}$ with a node set $\mathcal{S} = \{1, \cdots, NP\}$. The reproducing kernel (RK) approximation, $u^{RK}(\mathbf{x})$, of a function $u(\mathbf{x})$ is

$$u^{RK}(\mathbf{x}) = \sum_{I \in \mathcal{S}} \Psi_I(\mathbf{x}) d_I, \qquad (20)$$

with an RK shape function $\Psi_I(\mathbf{x})$ and a generalized nodal coefficient $d_I$. The RK shape function is a correction of a kernel function, $\Phi_a(\mathbf{x} - \mathbf{x}_I)$, defined on the compact support of node $I$ with a support size of $a$:

$$\Psi_I(\mathbf{x}) = C_I(\mathbf{x}) \Phi_a(\mathbf{x} - \mathbf{x}_I), \qquad (21)$$

where the kernel correction function $C_I(\mathbf{x})$ is defined as

$$C_I(\mathbf{x}) \equiv \left\{ \sum_{|\boldsymbol{\alpha}| \leq n} (\mathbf{x} - \mathbf{x}_I)^{\boldsymbol{\alpha}} b_{\boldsymbol{\alpha}}(\mathbf{x}) \right\}, \qquad (22)$$

where $(\mathbf{x} - \mathbf{x}_I)^{\boldsymbol{\alpha}}$ is a basis function, $\boldsymbol{\alpha} = (\alpha_1, \alpha_2, \ldots, \alpha_d)$ is a multi-dimensional index, and $|\boldsymbol{\alpha}| \equiv \sum_{i=1}^d \alpha_i$. $\mathbf{x}^{\boldsymbol{\alpha}}$ is defined as

$$\mathbf{x}^{\boldsymbol{\alpha}} \equiv x_1^{\alpha_1} \cdot x_2^{\alpha_2} \cdot \ldots \cdot x_d^{\alpha_d}. \qquad (23)$$

The coefficients, $b_{\boldsymbol{\alpha}}(\mathbf{x})$, are obtained by solving the following set of reproducing conditions:

$$\sum_{I \in \mathcal{S}} \Psi_I(\mathbf{x}) \mathbf{x}_I^{\boldsymbol{\alpha}} = \mathbf{x}^{\boldsymbol{\alpha}}, \qquad |\boldsymbol{\alpha}| \leq n. \qquad (24)$$

The results RK shape function takes the following explicit form:

$$\Psi_I(\mathbf{x}) = \mathbf{H}^T(\mathbf{0}) \mathbf{M}^{-1}(\mathbf{x}) \mathbf{H}(\mathbf{x} - \mathbf{x}_I) \Phi_a(\mathbf{x} - \mathbf{x}_I), \qquad (25)$$

where the moment matrix $\mathbf{M}(\mathbf{x})$ and the basis vector $\mathbf{H}(\mathbf{x} - \mathbf{x}_I)$ are defined as

$$\mathbf{M}(\mathbf{x}) = \sum_{I \in \mathcal{S}} \mathbf{H}(\mathbf{x} - \mathbf{x}_I) \mathbf{H}^T(\mathbf{x} - \mathbf{x}_I) \Phi_a(\mathbf{x} - \mathbf{x}_I), \qquad (26)$$

$$\mathbf{H}(\mathbf{x} - \mathbf{x}_I) = [1, (x_1 - x_{1I}), (x_2 - x_{2I}), (x_3 - x_{3I}), \cdots, (x_3 - x_{3I})^n]^T. \tag{27}$$

The kernel function $\Phi_a(\mathbf{x} - \mathbf{x}_I)$ determines the order of continuity, while the basis vector $\mathbf{H}(\mathbf{x} - \mathbf{x}_I)$ determines the polynomial completeness. Thus, it is straightforward to introduce high order continuity into the approximation space, independent of the basis order, which makes the RK approximation more appealing for approximating the smooth part of solution than the $C^0$ interpolation-type approximations used in finite element methods. Figure 1 shows a smooth RK shape function constructed on the linear basis.

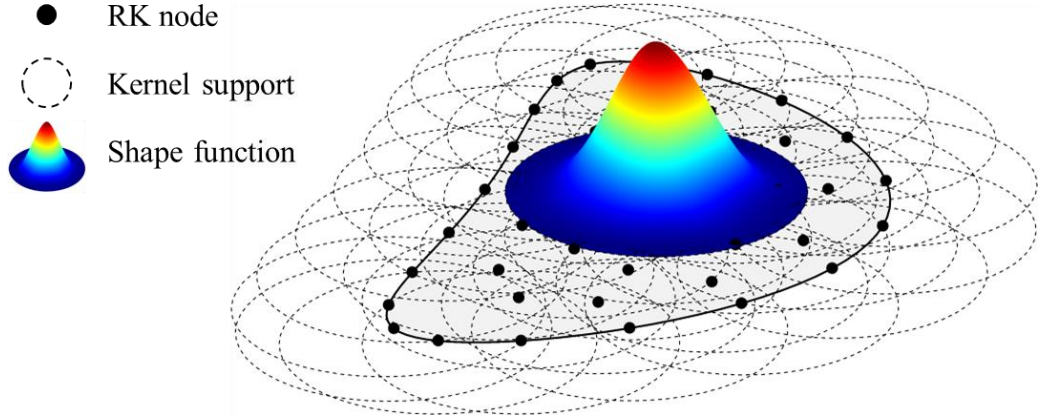

Figure 1. Illustration of RK discretization and shape function

For a quasi-uniform RK points distribution, the following global error estimation of standard RK approximation $u^{RK}$ holds, for $u \in H^r$, [41]

$$\|u^{RK} - u\|_{l,\Omega} \leq Cka^\gamma |u|_{p+1,\Omega}, \tag{28}$$

where $a$, $C$, $k$, $p$, and $\gamma = \min(p + 1 - l, \ r - l)$ are the support size, a generic constant, the number of overlapping points, the order of RK basis, and the convergence rate, respectively.

### 2.2.2. Stabilized conforming nodal integration

When Gauss integration (GI) is used for RKPM, a significantly high-order rule is required to yield optimal solution convergence, due to the rational shape function given in Eq. (26). This, in turn, leads to a significant increase in computational cost. To address this issue, the stabilized conforming nodal integration (SCNI) was proposed in [39]. SCNI enables optimal solution convergence for RKPM with a linear basis by satisfying the linear integration constraint. Compared to high-order GI, SCNI is computationally much more efficient as it eliminates the need to evaluate direct derivatives of RK shape functions at a large number of integration points. Additionally, Wei and Chen (2018) [40] show that the strain smoothing employed in SCNI helps to suppress spurious stress oscillation that can arise in localization problems. For this reason, SCNI is utilized to perform the domain integration required in Eq. (1).

In SCNI, the domain is partitioned into $N_{IC}$ conforming smoothing cells, such as Voronoi cells, as illustrated in Figure 2 where $N_{IC}$ denotes the number of smoothing cells. Note that, while $N_{IC}$ coincides with the number of particles for standard meshfree methods, the smoothing cells can be further refined to improve accuracy.

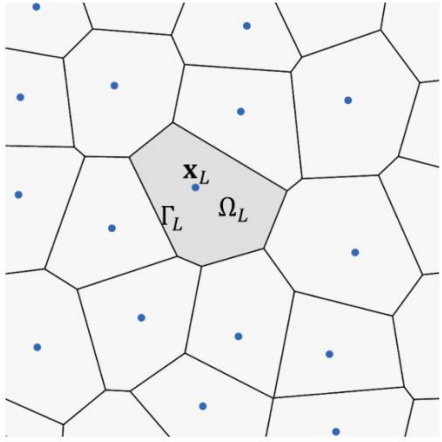

Figure 2. Conforming integration cell used in SCNI: $\Omega_L$, $\Gamma_L$, and $\mathbf{x}_L$ denote the domain, the boundary, and the centroid of the integration cell $L$, respectively.

The integration of the loss function by SCNI is performed as follows:

$$\int_\Omega \psi(\mathbf{u}^h, \nabla \mathbf{u}^h)\, d\Omega \approx \sum_L^{N_{IC}} \psi\left(\mathbf{u}^h(\mathbf{x}_L), \widetilde{\nabla}\mathbf{u}^h(\mathbf{x}_L)\right) V_L, \qquad (29)$$

where $\widetilde{\nabla}\mathbf{u}^h$ is the smoothed gradient of $\mathbf{u}$ defined as

$$\widetilde{\nabla}\mathbf{u}^h(\mathbf{x}_L) \equiv \frac{1}{V_L}\int_{\Gamma_L} \mathbf{u}^h(\mathbf{x}) \otimes \mathbf{n}(\mathbf{x})\, d\Gamma \approx \frac{1}{V_L}\sum_{k=1}^{N_{seg}^L} \mathbf{u}^h(\mathbf{x}_L^k) \otimes n_L^k, \qquad (30)$$

where $V_L$, $\mathbf{x}_L^k$, $n_L^k$, and $N_{seg}^L$ are the cell volume, the centroid of $k$-th boundary segment, the surface normal of $k$-th boundary segment, and the number of boundary segments of the integration cell $L$, respectively.

## 3. Neural Network-enhanced Reproducing Kernel Approximation

Figure 3 schematically illustrates a domain discretization by quasi-uniformly distributed background RK nodes, along with the evolving localizations in the domain. It is expected that the true solution would be rough near localizations and smooth in the remaining part of the domain. As discussed in Section 2.2.1, the RK approximation is intended to capture the smooth part of the solution. With the enrichment function (to be constructed) near the evolving localizations, the total solution is constructed by superposing a background RK approximation $u^{RK}(\mathbf{x})$ and a neural network (NN) approximation $u^{NN}(\mathbf{x})$ as follows: for $\mathbf{x} \in \Omega$,

$$u^h(\mathbf{x}) = u^{RK}(\mathbf{x}) + u^{NN}(\mathbf{x}), \tag{31}$$

where $u^h(\mathbf{x})$ is an NN-enhanced RK (NN-RK) approximation. With this construction, uniform RK discretization is considered as a background discretization, and the localized solution will be represented by the NN approximation. The NN-RK approximation utilizes the RK approximation's flexibility in selecting the order of continuity and the order of monomial bases.

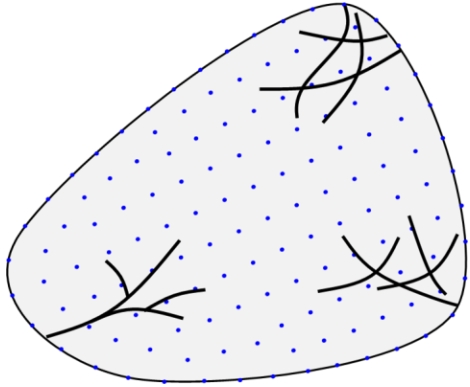

Figure 3. Schematic illustration of the NN-RK approximation: quasi-uniform background RK node distribution (blue dots) for smooth solution approximation and NN enrichment of the solution space for capturing localizations (black solid curves).

### 3.1.1. A neural network-based correction of RK approximation

In this section, we derive the NN-RK approximation through a neural network-based correction (NN-correction) of an RK approximation. Let $\Omega$ be a domain discretized by $NP$ background RK nodes with nodal coordinate $\{\mathbf{x}_I\}_{I \in \mathcal{S}}$ in a node set $\mathcal{S} = \{1, \cdots, NP\}$. In addition, define a node subset $\bar{\mathcal{S}}$ that contains the nodes with the associated RK shape functions to be corrected near localization. In this work, $\bar{\mathcal{S}} = \{J \mid \exists \mathbf{x} \in supp(\Psi_J), \psi_0^+(\mathbf{x}) \geq \kappa \psi_c\}$ with $\kappa = 0.5$ is applied. We start with the following NN-corrected RK approximation:

$$u^h(\mathbf{x}) = \sum_{I \in \mathcal{S}} \overline{\Psi}_I(\mathbf{x}) \bar{d}_I, \tag{32}$$

where the NN-corrected RK shape function $\overline{\Psi}_I(\mathbf{x})$ is defined as follows:

$$\overline{\Psi}_I(\mathbf{x}) = \begin{cases} \bar{C}_I(\mathbf{x})\Psi_I(\mathbf{x}), & I \in \bar{\mathcal{S}} \\ \Psi_I(\mathbf{x}), & I \in \mathcal{S}\setminus\bar{\mathcal{S}}, \end{cases} \qquad (33)$$

where $\Psi_I(\mathbf{x})$ and $\bar{C}_I(\mathbf{x})$ denote the original RK shape function defined in Section 2.2.1 and an NN-correction function, respectively. The NN-correction function takes the following form of a neural network with $n$ neurons possessed by the last hidden layer:

$$\bar{C}_I(\mathbf{x}) \equiv \bar{b}_I + \sum_{K=1}^{n} \bar{w}_{IK}\zeta_{IK}(\mathbf{x}), \qquad (34)$$

where $\bar{b}_I$, $\bar{w}_{IK}$, and $\zeta_{IK}(\mathbf{x})$ denote bias, weight, and last hidden layer's output. By substituting (33) and (34) into (32) and defining $d_I = \bar{b}_I \bar{d}_I$ and $w^C_{IK} = \bar{w}_{IK}\bar{d}_I$, we have a general expression of NN-RK approximation as follows:

$$u^h(\mathbf{x}) = u^{RK}(\mathbf{x}) + u^{NN}(\mathbf{x}), \qquad (35)$$

$$u^{RK} = \sum_{I \in \mathcal{S}} \Psi_I(\mathbf{x})d_I, \qquad (36)$$

$$u^{NN} = \sum_{I \in \bar{\mathcal{S}}} \sum_{K=1}^{n} \Psi_I(\mathbf{x})\zeta_{IK}(\mathbf{x})w^C_{IK}. \qquad (37)$$

***Remark 3.1.*** The background RK approximation $u^{RK}(\mathbf{x})$ in (36) is a standard RK approximation based on a polynomial RK basis. Meanwhile, the NN approximation $u^{NN}(\mathbf{x})$ in (37) contains nonstandard *adaptive* basis functions, which enables it to capture localized material responses with a coarse background RK discretization.

***Remark 3.2.*** As the RK shape functions possess the property of partition of unity, the NN-RK approximation

$$u^h(\mathbf{x}) = u^{RK}(\mathbf{x}) + u^{NN}(\mathbf{x}) = \sum_{I \in \mathcal{S}} \Psi_I(\mathbf{x})\left(d_I + \sum_{K=1}^{n} \zeta_{IK}(\mathbf{x})w^C_{IK}\right),$$
$$w^C_{IK} = 0, \qquad \forall I \in \mathcal{S}\setminus\bar{\mathcal{S}} \qquad (38)$$

can be viewed as patching the RK and NN approximations under the Partition of Unity framework.

***Remark 3.3.*** In (37), $\zeta_{IK}(\mathbf{x})$ is the activated output of $K$-th neuron in the last hidden layer of a neural network associated with node $I$. By having $\zeta_{IK}(\mathbf{x}) \equiv \zeta_K(\mathbf{x})$ for all $I \in \bar{\mathcal{S}}$, $\zeta_K(\mathbf{x})$ is detached from a specific background node and becomes a flexible *foreground* quantity. Then, the NN approximation in (37) can be rewritten as follows:

$$u^{NN} = \sum_{K=1}^{n} \zeta_K(\mathbf{x}) v_K(\mathbf{x}), \tag{39}$$

$$v_K(\mathbf{x}) \equiv \sum_{I \in \bar{S}} \Psi_I(\mathbf{x}) w_{IK}^C. \tag{40}$$

***Remark 3.4.*** The neural network to generate $\zeta_{IK}(\mathbf{x})$ can be either a traditional or a nonstandard neural network. In section 3.1.2, we present a modified deep neural network designed to effectively capture localizations.

### 3.1.2. Block-level neural network approximation

In this work, we introduce a modified deep neural network to increase the sparsity of the network architecture, improve the interpretability, and capture localizations effectively. In this regard, the following block-level NN approximation is introduced.

$$u^{NN} = \sum_{J=1}^{n_B} u_J^B(\mathbf{x}), \tag{41}$$

where $n_B$ is the number of NN blocks, and the block-level NN approximation $u_J^B(\mathbf{x})$ is defined as follows:

$$u_J^B(\mathbf{x}) = \sum_{K=1}^{n_{NK}} \hat{\phi}_{JK}(\mathbf{x}) \hat{v}_{JK}(\mathbf{x}), \tag{42}$$

$$\hat{v}_{JK}(\mathbf{x}) = \sum_{I \in \bar{S}} \Psi_I(\mathbf{x}) \hat{w}_{IJK}^C, \tag{43}$$

where $\hat{\phi}_{JK}(\mathbf{x})$ and $n_{NK}$ are $K$-th NN kernel function in $J$-th NN block and the number of NN kernel functions per NN block, respectively. Note that (41)-(43) are shown to be equivalent to (39) and (40) by flattening the indices $JK$ in (42) and (43) into $K$.

Figure 4 illustrates the modified network architecture of $J$-th NN block, for which the construction is made so that the neural network approximation can capture complicated localization topologies effectively. Also, the construction of the neural network at the block level significantly increases the sparsity of the weight matrices, compared to the densely connected standard deep neural networks utilized in many previous studies in literature [29,34]. As shown in Figure 4, three sets of unknown parameters are involved in the NN approximation: the location-control weight set $\mathbf{W}_J^L$, the shape-control weight set $\mathbf{W}_J^S$ as well as the NN-correction weight set $\mathbf{W}_J^C = \left\{\{\hat{w}_{IJK}^C\}_{I \in \bar{S}}\right\}_{K=1}^{n_{NK}}$ in (43). These parameters are to be automatically determined by solving the minimization problem (1). Details on the sub-blocks described in Figure 4 and their associated unknown parameters are explained in the following subsections.

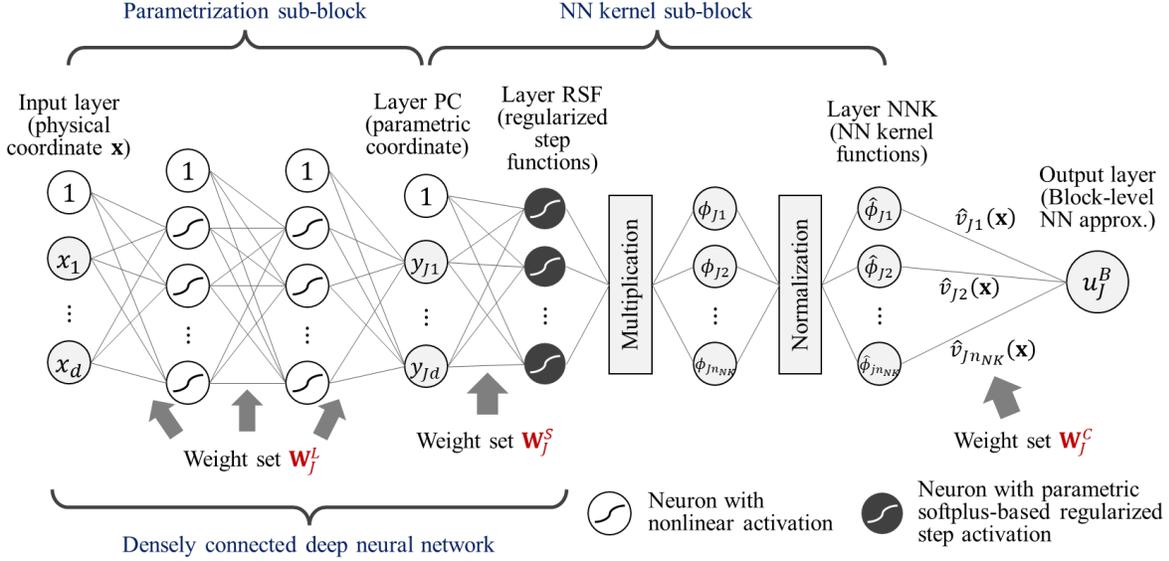

Figure 4. Modified neural network architecture of $J$-th NN block. The unknown parameters introduced in each part are denoted in red color.

### 3.1.3. Parametrization sub-block

As shown in Figure 4, the parametric coordinate $\mathbf{y}_J$ in Layer PC is the output of the parametrization sub-block, which is an intermediate variable of a densely connected deep neural network $\mathcal{N}: \mathbf{x} \to \mathbf{y}_J$ that takes $\mathbf{x} \in \mathbb{R}^d$ and $\mathbf{y}_J \equiv \mathbf{y}(\mathbf{x}; \mathbf{W}_J^L) \in \mathbb{R}^d$ as its input and output, respectively. The parametrization projects complicated localization patterns onto a parametric space, so that complicated localizations can be captured with NN kernel functions in a simple mathematical form. With $n_{HL}$ hidden layers, the function $\mathbf{y}(\mathbf{x}; \mathbf{W}_J^L)$ is defined as

$$\mathbf{y}(\mathbf{x}; \mathbf{W}_J^L) = \mathbf{f}\left(\cdot; \{\mathbf{w}_{J(n_{HL}+1)}^L, b_{J(n_{HL}+1)}^L\}\right) \circ \mathbf{h}\left(\cdot; \{\mathbf{w}_{Jn_{HL}}^L, b_{Jn_{HL}}^L\}\right) \circ \cdots \circ \mathbf{h}\left(\mathbf{x}; \{\mathbf{w}_{J1}^L, b_{J1}^L\}\right) \quad (44)$$

with

$$\mathbf{h}(\boldsymbol{\xi}; \{\mathbf{w}_{Jl}^L, b_{Jl}^L\}) = a\left(\mathbf{f}(\boldsymbol{\xi}; \{\mathbf{w}_{Jl}^L, b_{Jl}^L\})\right), \quad (45)$$

$$\mathbf{f}(\boldsymbol{\xi}; \{\mathbf{w}_{Jl}^L, b_{Jl}^L\}) = \mathbf{w}_{Jl}^L \boldsymbol{\xi} + b_{Jl}^L. \quad (46)$$

In (44), $\mathbf{w}_{Jl}^L$ and $b_{Jl}^L$ denote weight and bias of layer $l$, respectively, and the location-control parameter set $\mathbf{W}_J^L$ in Figure 4 is defined as $\mathbf{W}_J^L = \{\mathbf{w}_{Jl}^L, b_{Jl}^L\}_{l=1}^{n_{HL}+1}$. In (45), $a(\cdot)$ denote an activation function. In this work, the hyperbolic tangent activation function is used.

### 3.1.4. NN kernel function

As shown in Figure 4, the NN kernel functions $\hat{\phi}_{JK}(\mathbf{x})$ in Layer NNK is the outcome of the normalization of unnormalized NN kernel functions $\phi_{JK}(\mathbf{x})$. The normalization is defined as

$$\hat{\phi}_{JK}(\mathbf{x}) = \frac{\phi_{JK}(\mathbf{x})}{\sum_{I=1}^{n_B} \sum_{L=1}^{n_{NK}} \phi_{IL}(\mathbf{x})}, \tag{47}$$

and the NN kernel function $\phi_{JK}(\mathbf{x})$ is defined as

$$\phi_{JK}(\mathbf{x}) = \prod_{\alpha=1}^{d} \prod_{i=1}^{2} \bar{\phi}_i\left(y_{J\alpha}; \{\bar{y}_{\alpha i}^{JK}, c_{\alpha i}^{JK}, \beta_{\alpha i}^{JK}\}\right), \tag{48}$$

where $\bar{\phi}_i$ and $\{\bar{y}_{\alpha i}^{JK}, c_{\alpha i}^{JK}, \beta_{\alpha i}^{JK}\}$ denote a regularized step function and shape-control parameters, respectively. The shape-control weight set $\mathbf{W}_J^S$ in Figure 4 is defined as $\mathbf{W}_J^S = \left\{\left\{\{\bar{y}_{\alpha i}^{JK}, c_{\alpha i}^{JK}, \beta_{\alpha i}^{JK}\}_{\alpha=1}^{d}\right\}_{i=1}^{2}\right\}_{K=1}^{n_{NK}}$. In this work, the regularized step function is constructed based on the parametric softplus activation function $S$ defined as follows:

$$\bar{\phi}_i(y; \{\bar{y}_i, c_i, \beta_i\}) = S\left(z_i(y) + \frac{1}{2}; \beta_i\right) - S\left(z_i(y) - \frac{1}{2}; \beta_i\right), \tag{49}$$

$$z_i(y) = (-1)^i (y - \bar{y})/c, \quad i = 1, 2, \tag{50}$$

$$S(z; \beta) = \frac{1}{\beta} \log(1 + e^{\beta z}). \tag{51}$$

In (49)-(51), $\beta_i$ controls the sharpness in the transition of derivative as shown in Figure 5 (a-b), and $c_i$ controls the sharpness of the solution transition as shown in Figure 5 (c). In addition, $\bar{y}_i$ influences the support of $\bar{\phi}_i$. Note that $\bar{\phi}_i$ is the output of Layer RSF in Figure 4, and $(1/c_i)$ and $(-\bar{y}_i/c_i)$ are respectively the weight and the bias of Layer RSF. Figure 6 shows a schematic illustration of a two-dimensional NN kernel which possesses a sharp transition in direction $y$. Interested readers refer to [33] for more details on the NN kernel functions.

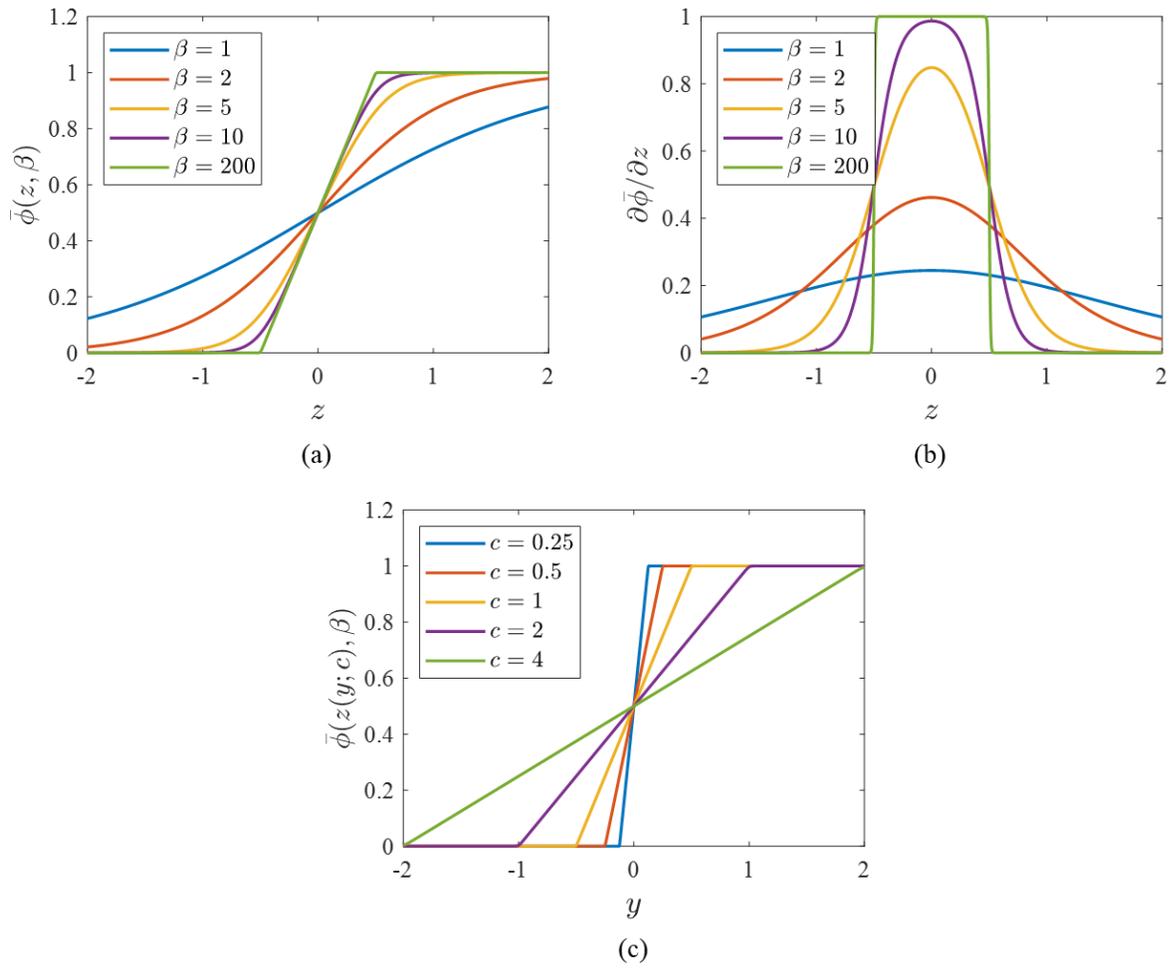

Figure 5. The influence of the control parameters on solution transition: (a) the influence of $\beta$ on $\bar{\phi}$, (b) the influence of $\beta$ on $\partial\bar{\phi}/\partial z$, and (c) the influence of $c$ on $\bar{\phi}$ with $\beta = 200$

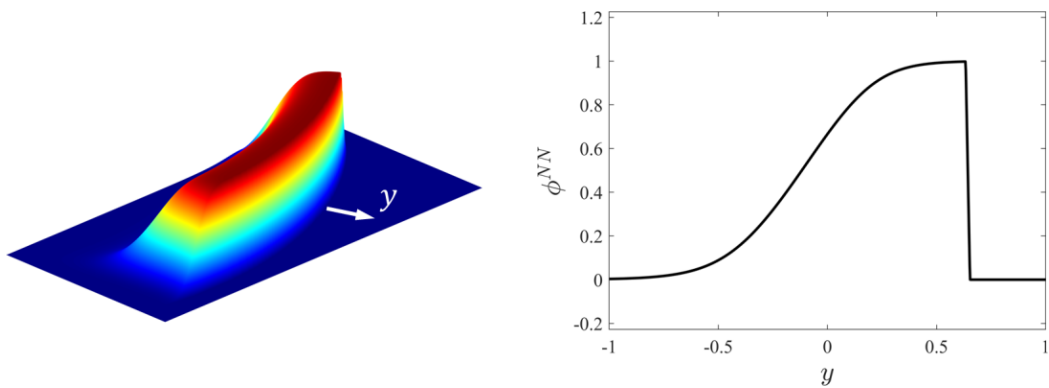

Figure 6. Schematic illustration of an NN kernel function: (left) two-dimensional NN kernel function $\phi$ and (right) its cross-sectional value across $y$.

## 3.2. Convergence Properties

An error bound of the proposed NN-RK approximation is estimated. Let $\widehat{\Omega}$ be the transition zone near the localization domain. Then, we have

$$\|u^h - u\|_{0,\Omega} \leq \|u^h - u\|_{0,\Omega\setminus\widehat{\Omega}} + \|u^h - u\|_{0,\widehat{\Omega}}. \tag{52}$$

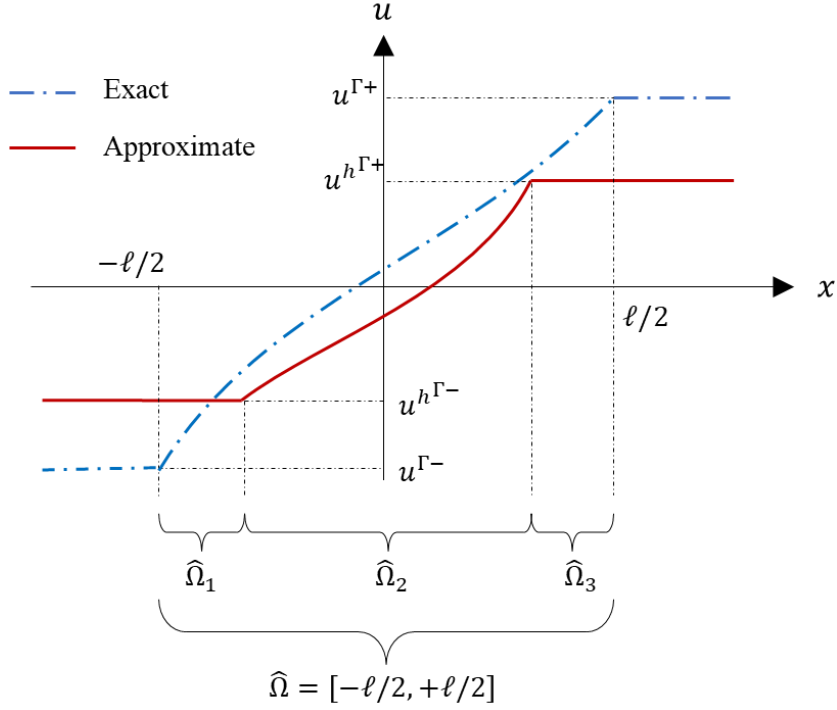

Figure 7. Arbitrary $u$ with a sharp transition occurring in the transition zone $\widehat{\Omega}$ and its approximation $u^h$ with a sharp transition occurring in $\widehat{\Omega}_2$

As shown in Figure 7, we consider an arbitrary $u$ with a sharp transition occurring in $\widehat{\Omega} = [-\ell/2, +\ell/2]$ and its approximation $u^h$ with a transition occurring in $\widehat{\Omega}_2$. For both $u$ and $u^h$, it is assumed that there are weak discontinuities on the boundaries of the transition zones. For brevity, let us introduce the following function $w$:

$$w(\chi; \xi) \equiv \frac{[\![\xi]\!]}{\ell}\chi + \langle\!\langle\xi\rangle\!\rangle, \tag{53}$$

where $[\![\xi]\!] \equiv \xi^+ - \xi^-$ and $\langle\!\langle\xi\rangle\!\rangle \equiv (\xi^+ + \xi^-)/2$ are a difference operator and an average operator, respectively, with $\xi^+ \equiv \xi(x = +\ell/2)$ and $\xi^- \equiv \xi(x = -\ell/2)$. Using (53), the true solution $u$ in the transition domain $\widehat{\Omega}$ can be written in a parametric coordinate $y^u$ as

$$u(x) = w(y^u(x); u^\Gamma), \tag{54}$$

where $u^\Gamma$ is the value of $u$ on the boundary of $\widehat{\Omega}$, and, from (53) and (54), $y^u(x)$ is obtained as

$$y^u(x) = \frac{\ell}{[\![u^\Gamma]\!]}(u(x) - \langle\!\langle u^\Gamma\rangle\!\rangle). \tag{55}$$

Similarly, the approximated solution $u^h(x)$ in the transition domain $\widehat{\Omega}$ is written in an approximated parametric coordinate $Y$ as

$$u^h(x) = w\left(Y(x); u^{h\Gamma}\right), \tag{56}$$

with

$$Y(x) = \begin{cases} -\ell/2, & x \in \widehat{\Omega}_1 \\ y(x), & x \in \widehat{\Omega}_2, \\ \ell/2, & x \in \widehat{\Omega}_3 \end{cases} \tag{57}$$

where $y(x)$ is the neural network-based parametrization defined in (44), and $u^{h\Gamma}$ is the value of $u^h$ on the boundary of $\widehat{\Omega}$. In (57), the subdomains are defined as $\widehat{\Omega}_1 = \{x \mid y(-\ell/2) \leq y(x) \leq -\ell/2\}$, $\widehat{\Omega}_2 = \{x \mid -\ell/2 < y(x) \leq \ell/2\}$, and $\widehat{\Omega}_3 = \{x \mid \ell/2 < y(x) \leq y(\ell/2)\}$. Note that, with $\beta \to \infty$, the NN kernel function defined in (48)-(50) introduces weak discontinuities on $y(x) = \pm\ell/2$.

With (54) and (56), the last term in (52) becomes

$$\begin{aligned}
\|u^h - u\|_{0,\widehat{\Omega}} &= \left\|w\left(Y(x); u^{h\Gamma}\right) - w(y^u(x); u^\Gamma)\right\|_{0,\widehat{\Omega}} \\
&= \left\|w\left(Y(x); u^{h\Gamma}\right) - w(Y(x); u^\Gamma) + w(Y(x); u^\Gamma) - w(y^u(x); u^\Gamma)\right\|_{0,\widehat{\Omega}} \\
&= \left\|w\left(Y(x); u^{h\Gamma} - u^\Gamma\right) + w(Y(x); u^\Gamma) - w(y^u(x); u^\Gamma)\right\|_{0,\widehat{\Omega}} \\
&\leq \left\|w\left(Y(x); u^{h\Gamma} - u^\Gamma\right)\right\|_{0,\widehat{\Omega}} + \left\|w(Y(x); u^\Gamma) - w(y^u(x); u^\Gamma)\right\|_{0,\widehat{\Omega}}.
\end{aligned} \tag{58}$$

The first term on the right-hand side of (58) is bounded as follows:

$$\begin{aligned}
\left\|w\left(Y(x); u^{h^\Gamma} - u^\Gamma\right)\right\|_{0,\widehat{\Omega}} &= \left\|\left(\llbracket u^{h^\Gamma} - u^\Gamma \rrbracket / \ell\right) Y(x) + \langle\!\langle u^{h^\Gamma} - u^\Gamma\rangle\!\rangle\right\|_{0,\widehat{\Omega}} \\
&\leq \left\|\left|u^{h^{\Gamma-}} - u^{\Gamma-}\right| + \left|u^{h^{\Gamma+}} - u^{\Gamma+}\right|\right\|_{0,\widehat{\Omega}} \\
&\leq \left\|u^{h^{\Gamma-}} - u^{\Gamma-}\right\|_{0,\widehat{\Omega}} + \left\|u^{h^{\Gamma+}} - u^{\Gamma+}\right\|_{0,\widehat{\Omega}} \\
&= \ell^{1/2}\left(\left|u^{h^{\Gamma-}} - u^{\Gamma-}\right| + \left|u^{h^{\Gamma+}} - u^{\Gamma+}\right|\right)
\end{aligned} \quad (59)$$

The second term on the right-hand side of (58) is bounded as follows:

$$\begin{aligned}
\left\|w(Y(x); u^\Gamma) - w(y(x); u^\Gamma)\right\|_{0,\widehat{\Omega}} &= \left\|\frac{\llbracket u^\Gamma \rrbracket}{\ell}(Y(x) - y^u(x))\right\|_{0,\widehat{\Omega}} \\
&= \frac{|\llbracket u^\Gamma \rrbracket|}{\ell}\|Y(x) - y^u(x)\|_{0,\widehat{\Omega}} \\
&\leq \frac{|\llbracket u^\Gamma \rrbracket|}{\ell}\|y(x) - y^u(x)\|_{0,\widehat{\Omega}}.
\end{aligned} \quad (60)$$

Therefore, for $\widehat{\Omega}$, the following error bound is obtained.

$$\|u^h - u\|_{0,\widehat{\Omega}} \leq \ell^{1/2}\left(\left|u^{h^{\Gamma-}} - u^{\Gamma-}\right| + \left|u^{h^{\Gamma+}} - u^{\Gamma+}\right|\right) + \frac{|\llbracket u^\Gamma \rrbracket|}{\ell}\|y(x) - y^u(x)\|_{0,\widehat{\Omega}}. \quad (61)$$

For multi-dimensions, we have

$$\|u^h - u\|_{0,\widehat{\Omega}} \leq \ell^{1/2}\|u^h - u\|_{0,\widehat{\Gamma}} + \frac{|\llbracket u^\Gamma \rrbracket|}{\ell}\|y(x) - y^u(x)\|_{0,\widehat{\Omega}}, \quad (62)$$

where $\widehat{\Gamma} \equiv \partial\widehat{\Omega}\backslash\partial\Omega$ denotes the interface of weak discontinuity. Using the Sobolev trace inequality and (28), the first term on the right-hand side of (62) is bounded as follows: with a generic constant $\hat{C}$ and $\hat{\hat{C}}$,

$$\begin{aligned}
\|u^h - u\|_{0,\widehat{\Gamma}} &\leq \|u^h - u\|_{0,\partial(\Omega\backslash\widehat{\Omega})} \leq \hat{C}\|u^h - u\|_{0,\Omega\backslash\widehat{\Omega}}^{1/2}\|u^h - u\|_{1,\Omega\backslash\widehat{\Omega}}^{1/2} \\
&\leq \hat{\hat{C}} k a^{\hat{\gamma}} |u|_{p+1,\Omega\backslash\widehat{\Omega}},
\end{aligned} \quad (63)$$

where $\hat{\gamma} = \max(p + 0.5, \tilde{r})$ where $\tilde{r}$ and $p$ denotes the regularity of $u$ in $\Omega\backslash\widehat{\Omega}$ and the order of basis of the background RK discretization, respectively. With (28), (62), and (63), the global error (52) has the following error bound:

$$\|u^h - u\|_{0,\Omega} \leq \left(Ca^\gamma + \hat{\hat{C}}a^{\hat{\gamma}}\right) k|u|_{p+1,\Omega\backslash\widehat{\Omega}} + \frac{|\llbracket u^\Gamma \rrbracket|}{\ell}\|y(x) - y^u(x)\|_{0,\widehat{\Omega}}, \quad (64)$$

where $\gamma = \max(p + 1, \tilde{r})$. For smooth $u$ in $\Omega\backslash\widehat{\Omega}$, $\hat{\gamma} = \gamma - 0.5$ holds, which means that $\hat{\gamma}$ dominates the first term on the right-hand side of (64), leading to

$$\|u^h - u\|_{0,\Omega} \leq \left(C + \hat{C}\right) a^{\hat{\gamma}} k |u|_{p+1,\Omega\backslash\hat{\Omega}} + \frac{|[\![u^\Gamma]\!]|}{\ell} \|y(x) - y^u(x)\|_{0,\hat{\Omega}}, \quad (65)$$

In the last term of (65), $\|y(x) - y^u(x)\|_{0,\hat{\Omega}}$ denotes the parametrization error. (65) implies that, when the parametrization error is relatively large, the solution convergence will be governed by the convergence of the parametrization. Conversely, for $\|y(x) - y^u(x)\|_{0,\Gamma} \to 0$, the convergence will be governed by the background RK discretization with a rate of $\hat{\gamma}$, e.g., 1.5 when a linear RK basis is used. The error bound of $\|y(x) - y^u(x)\|_{0,\Gamma}$ follows the universal approximation theorem [1,37] when a neural network is used for parametrization. For example, for a neural network with a single hidden layer, the error bound is estimated as follows [1]: with a generic constant $C_y < \infty$,

$$\|y(x) - y^u(x)\|_{0,\Gamma} \leq C_y n_{NR}^{-1/2}, \quad (66)$$

which leads to the following error estimation of NN-RK approximation

$$\|u^h - u\|_{0,\Omega} \leq \left(C + \hat{C}\right) a^{\hat{\gamma}} k |u|_{p+1,\Omega\backslash\hat{\Omega}} + C_y \frac{|[\![u^\Gamma]\!]|}{\ell} n_{NR}^{-1/2}. \quad (67)$$

### 3.3. Regularization

To avoid the potential loss of ellipticity of the problem and the resulting discretization sensitivity in the numerical solution of the local problem defined in Section 2, a regularization treatment is needed. A straightforward remedy is to impose a proper constraint such that the physical bandwidth of the damage does not become narrower than a certain limit. To analyze a localization width possessed by the NN-RK approximation, we start with a Taylor expansion of the parametric coordinate as follows:

$$y(\mathbf{x}) \approx \bar{y} + (\mathbf{x} - \bar{\mathbf{x}}) \cdot \nabla^x y(\bar{\mathbf{x}}), \quad (68)$$

where $\bar{y} = y(\bar{\mathbf{x}})$, and $\bar{y}$ is defined in Section 3.1.4, for which the superscripts and subscripts are omitted for brevity. With (68), $z$ defined in (50) is written as

$$z(y(\mathbf{x}); \{\bar{y}, c\}) = \frac{(y(\mathbf{x}) - \bar{y})}{c} \approx \frac{(\mathbf{x} - \bar{\mathbf{x}}) \cdot \nabla^x y(\bar{\mathbf{x}})}{c} \equiv \frac{\bar{\xi}(\mathbf{x}; \bar{\mathbf{x}})}{c}, \quad (69)$$

with $\bar{\xi}(\mathbf{x}; \bar{\mathbf{x}}) \equiv (\mathbf{x} - \bar{\mathbf{x}}) \cdot \nabla^x y(\bar{\mathbf{x}})$. When $\|\nabla^x y(\bar{\mathbf{x}})\| = 1$, $\bar{\xi}(\mathbf{x}; \bar{\mathbf{x}})$ in (69) is a projection of the physical coordinate onto the direction normal to the localization. Therefore, by satisfying conditions

$$\|\nabla^x y(\bar{\mathbf{x}})\| \leq 1, \tag{70}$$
$$c \geq \ell,$$

the transition width of $\bar{\phi}$ in (49) has a lower bound of $\ell$, and thus the localization width in the NN-RK approximation has the same lower bound. In this work, a constraint $\|\nabla^x y\| \leq 1$ is imposed in the loss function (1), and the lower bound of the sharpness control parameter $c$ in (50) is set to an NN length scale parameter $\ell$. The modified loss function with regularization reads:

$$\min_{\mathbf{u}} \overline{\Pi}(\mathbf{u}, \mathbf{y}) = \Pi(\mathbf{u}) + \Pi^{\text{Reg}}(\mathbf{y}),$$
$$\Pi^{\text{Reg}}(\mathbf{y}) = \frac{\kappa\mu}{2} \sum_{\alpha,J} \int_{\Omega} \langle \|\nabla^x y_{J\alpha}(\mathbf{x})\| - 1 \rangle_+^2 \, d\Omega, \tag{71}$$

where $\Pi$ is the potential function defined in (1), and $\kappa$ is the normalized penalty parameter. In this work, $\kappa = 10^4$ is used. Note that this approach is different from the $\widehat{H}$-regularization introduced by Baek et al. (2022) [33] in which the parametric coordinates are directly scaled by $\widehat{H}$ as follows:

$$z = \frac{(y - \bar{y})\widehat{H}}{c}, \qquad \text{where } \widehat{H} \equiv 1/\max(\|\nabla^x y\|, 1). \tag{72}$$

An advantage of the regularization designed in this work over the $\widehat{H}$-regularization is that the necessity to compute the second order gradient of $y$ for the evaluation of the strain energy in the loss function is avoided.

## 4. Numerical implementation

The minimization problem is rewritten as follows:

$$\min_{\mathbf{d}, \mathbf{W}} \left[ \Pi\big(\mathbf{u}^h(\mathbf{d}, \mathbf{W})\big) + \Pi^{\text{Reg}}\big(\mathbf{y}(\mathbf{x}; \mathbf{W}^L)\big) \right], \tag{73}$$

where $\mathbf{u}^h(\mathbf{d}, \mathbf{W}) = \mathbf{u}^{RK}(\mathbf{d}) + \mathbf{u}^{NN}(\mathbf{W})$ is the NN-RK approximation with the RK coefficient set, $\mathbf{d}$, and the neural network weight set, $\mathbf{W} = \{\mathbf{W}^L, \mathbf{W}^S, \mathbf{W}^C\}$ with $\mathbf{W}^L = \{\mathbf{W}^L_J\}_{J=1}^{n_B}$, $\mathbf{W}^S = \{\mathbf{W}^S_J\}_{J=1}^{n_B}$, and $\mathbf{W}^C = \{\mathbf{W}^C_J\}_{J=1}^{n_B}$. In (73), $\psi$ and $F$ denote the energy density and the external work defined in (1), respectively.

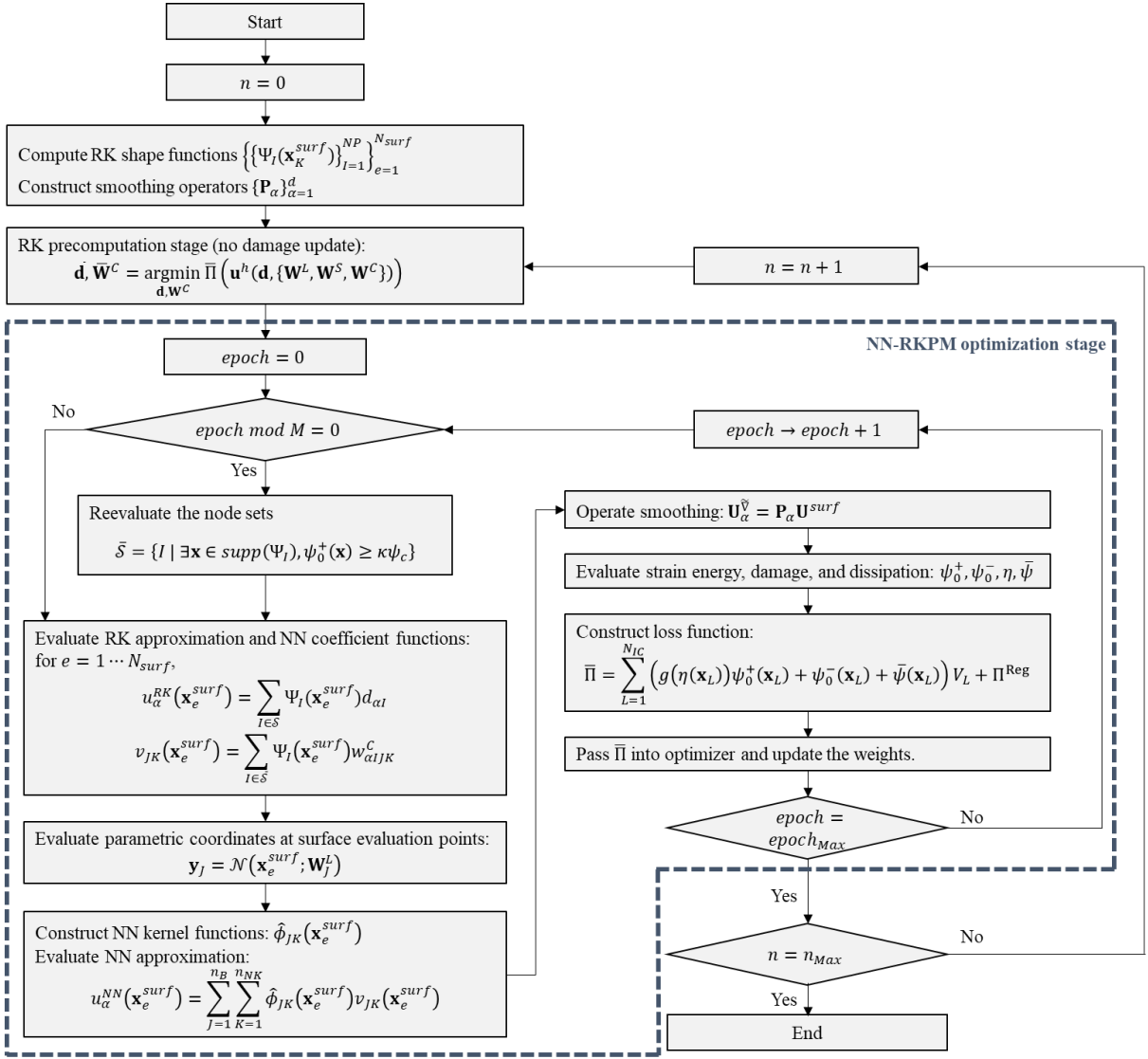

Figure 8. Flowchart of the solution procedure

Figure 8 shows the flowchart of the solution procedure. In the flowchart $n$ and $n_{Max}$ denotes the loading step and the maximum loading step, respectively. At loading step $n + 1$, the solution procedure mainly consists of two parts: RK precomputation stage and NN-RKPM optimization stage.

A. <u>RK precomputation stage</u>

To obtain the initial guesses $\bar{\mathbf{d}}^{(n+1)}$ and $\bar{\mathbf{W}}^{C(n+1)}$ to be used in the NN-RKPM optimization stage, the minimization problem (73) is first solved only for $\mathbf{d}^{(n+1)}$ and $\mathbf{W}^{C(n+1)}$:

$$\bar{\mathbf{d}}^{(n+1)}, \mathbf{W}^{C\,(n+1)} = \underset{\mathbf{d}, \mathbf{W}^C}{\operatorname{argmin}} \left[ \Pi\left(\mathbf{u}^h\left(\mathbf{d}, \{\mathbf{W}^{L(n)}, \mathbf{W}^{S(n)}, \mathbf{W}^C\}\right)\right) + \Pi^{\text{Reg}}\left(\mathbf{y}\left(\mathbf{x}; \mathbf{W}^{L(n)}\right)\right) \right] \quad (74)$$

$$\text{subjected to } \mathbf{u}(\mathbf{x}) = \mathbf{g}^{(n+1)} \text{ on } \partial\Omega_g.$$

In this stage, the weight sets $\{\mathbf{W}^L, \mathbf{W}^S\}$ and the damage $\eta$ from the previous loading step are used. Also, the damage is not updated. This is equivalent to the standard Galerkin-based RKPM problem and can be solved by a standard matrix solver.

B. <u>NN-RKPM optimization stage</u>

In the second stage, the minimization problem (73) is solved for the entire unknown parameters $\mathbf{d}$ and $\mathbf{W}$.

$$\bar{\mathbf{d}}^{(n+1)}, \mathbf{W}^{(n+1)} = \underset{\mathbf{d}, \mathbf{W}}{\operatorname{argmin}} \left[ \Pi\left(\mathbf{u}^h(\mathbf{d}, \mathbf{W})\right) + \Pi^{\text{Reg}}(\mathbf{y}(\mathbf{x}; \mathbf{W}^L)) \right] \quad (75)$$

$$\text{subjected to } \mathbf{u}(\mathbf{x}) = \mathbf{g}^{(n+1)} \text{ on } \partial\Omega_g.$$

In this stage, the damage is updated as well. The minimization problem can be solved iteratively by a suitable optimizer. In this work, *Adam* [42], a first-order optimizer with adaptive learning rate, is used for the first several epochs. Then, the optimizer is switched to limited-memory Broyden–Fletcher–Goldfarb–Shanno algorithm (L-BFGS) [43], a second-order optimizer, for the remaining optimization.

For domain integration involved in (73), SCNI introduced in Section 2.2.2 is used with refined smoothing cells near localization. As discussed in Section 2.2.2, the advantage of using SCNI for the proposed method is twofold: 1) it eliminates the requirement of computing the computationally expensive direct derivative of $\mathbf{u}^{NN}$ with the automatic differentiation to evaluate strain and stress, and 2) it suppresses stress oscillations.

For computationally efficient implementation of the strain smoothing operation in SCNI, precomputed sparse smoothing matrices $\mathbf{P}_\alpha$ with $\alpha = 1 \cdots d$ can be considered to perform the following global smoothing:

$$\mathbf{U}_\alpha^{\widetilde{\nabla}} = \mathbf{P}_\alpha \mathbf{U}^{surf}, \quad (76)$$

by which the strain smoothing in all the smoothing cells as discussed in section 2 are conducted simultaneously. In (76), $\mathbf{U}_\alpha^{\widetilde{\nabla}} = \left[\widetilde{u_{,\alpha}^h}(\mathbf{x}_1), \cdots, \widetilde{u_{,\alpha}^h}(\mathbf{x}_L), \cdots, \widetilde{u_{,\alpha}^h}(\mathbf{x}_{N_{IC}})\right]^T$ is a column vector containing the smoothed gradients of $u^h$ with respect to $x_\alpha$ for all the smoothing cells in the domain, i.e., $L = 1 \cdots N_{IC}$. $\mathbf{U}^{surf} = \left[u^h(\mathbf{x}_1^{surf}), \cdots, u^h(\mathbf{x}_e^{surf}), \cdots, u^h\left(\mathbf{x}_{N_{seg}}^{surf}\right)\right]^T$ is a column vector containing $u^h$ evaluated at a smoothing cell surface evaluation point $\mathbf{x}_e^{surf}$ for $e = 1 \cdots N_{surf}$, where $N_{surf}$ denotes the total number of smoothing cell surface evaluation points in the domain. The $(L, e)$ component of the smoothing operator $\mathbf{P}_\alpha$ is

$$P_{\alpha Le} = \begin{cases} \dfrac{1}{V_L} A_e n_\alpha^K, & \text{if } \Gamma_e \subset \Omega_L \\ 0, & \text{otherwise} \end{cases}, \tag{77}$$

where $\Gamma_e$, $\Omega_L$, $n_\alpha^K$, and $A_e$ denote $e$-th smoothing cell surface segment, $L$-th smoothing cell domain, $\alpha$-th component of the surface normal, and the area of $e$-th smoothing cell surface segment, respectively. The same procedure can be used to compute $\widetilde{\nabla} y_i$ for Eq. (71).

## 5. Numerical Examples

Several numerical examples are presented to demonstrate the proposed method's accuracy, regularization ability, and capability to capture complicated localization patterns. Unless otherwise specified, for the RK approximation, the linear basis with cubic B-spline kernel function of normalized support size 2.0 is used, and, for the NN approximation, a single 4-kernel NN block is used along with a densely connected neural network with the hyperbolic tangent activation function for the parametrization sub-block. For the domain integration, SCNI is used with refined smoothing cells in the zone along the expected damage path.

### 5.1. Elasticity with pre-existing damaged zone

Consider a domain $[-L/2, L/2] \times [-H/2, H/2]$ with a degraded zone with width $w$. We consider two different cases of pre-existing damaged zone geometry, as show in Figure 9(a) and (b). For both cases, $L = 2$ mm and $H = 0.5$ mm are used. For Case I, the degraded zone is vertically aligned at the center of the domain. For Case II, the anti-symmetric degraded zone is centered at the origin with $\mathbf{x}_{c1} = (-0.1, -0.5)$, $R_1 = 0.35$, $\mathbf{x}_{c2} = (-0.1, 0)$, and $R_2 = 0.1$ in unit of mm. For both cases, Dirichlet boundary conditions are applied to the left and right surfaces with $g = 1 \times 10^{-2}$ mm, and zero traction boundary conditions are applied to the top and bottom surfaces. For Case I, $w = H/100$, $E = 210$ GPa, and $\nu = 0$ are used, and for Case II, $w = H/1000$, $E = 210$ GPa, and $\nu = 0.3$ are used. The Young's modulus within the degraded zones is $kE$ with $k = 10^{-2}$ for Case I and $k = 10^{-3}$ for Case II.

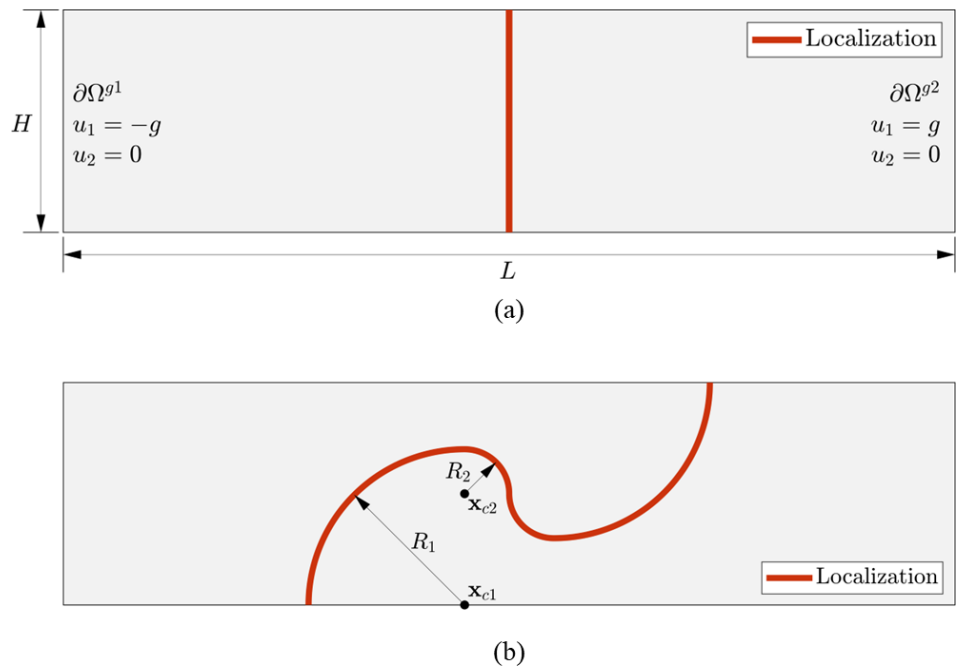

Figure 9. Geometry and boundary conditions for problem of elasticity with pre-existing damaged zone: (a) Case I and (b) Case II

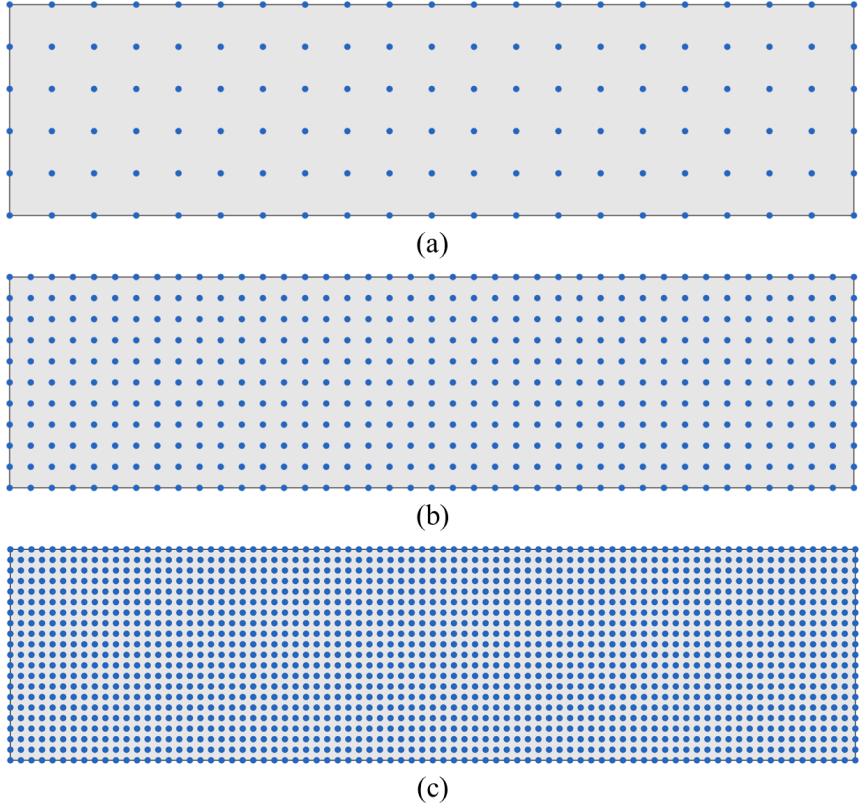

Figure 10. Background RK discretizations used for the elasticity with pre-damaged material: (a) $21 \times 6$ RK nodes with $h = H/5$, (b) $41 \times 11$ RK nodes with $h = H/10$, and (c) $81 \times 21$ RK nodes with $h = H/20$

For Case I, the exact solution is as follows:

$$u_1(\mathbf{x}) = \begin{cases} b(x_1 + L) - g, & x_1 \leq -w/2 \\ (b/k)x_1, & -w/2 < x_1 \leq w/2 \\ b(x_1 - L) + g, & x_1 > w/2 \end{cases} \quad (78)$$

$$u_2(\mathbf{x}) = 0$$

where $b = 2g/\big((1/k - 1)w + 2L\big)$. For the numerical solution, the domain is uniformly discretized by $21 \times 6$ RK nodes (see Figure 10 (a)), and a single 10-neuron hidden layer is used for the parametrization sub-block. Figure 11 shows the displacement predicted by the proposed method. The numerical solution captures the sharp transition in the horizontal displacement very well along with the zero vertical displacement due to zero Poisson's ratio. As shown in the figures in the 2nd row in Figure 11, the NN approximation appears near the localization capturing the sharp transition of $u_1$, and the RK approximation captures the solution in the other area, with smooth transition between two approximations. Figure 12 shows the horizontal displacement and normal strain along $y = 0$ in which the numerical solution is shown to be highly accurate compared to the exact solution. The computed $L_2$ norm and $H^1$ semi-norm of the solution error are $2.921 \times 10^{-4}$ and $2.437 \times 10^{-6}$, respectively.

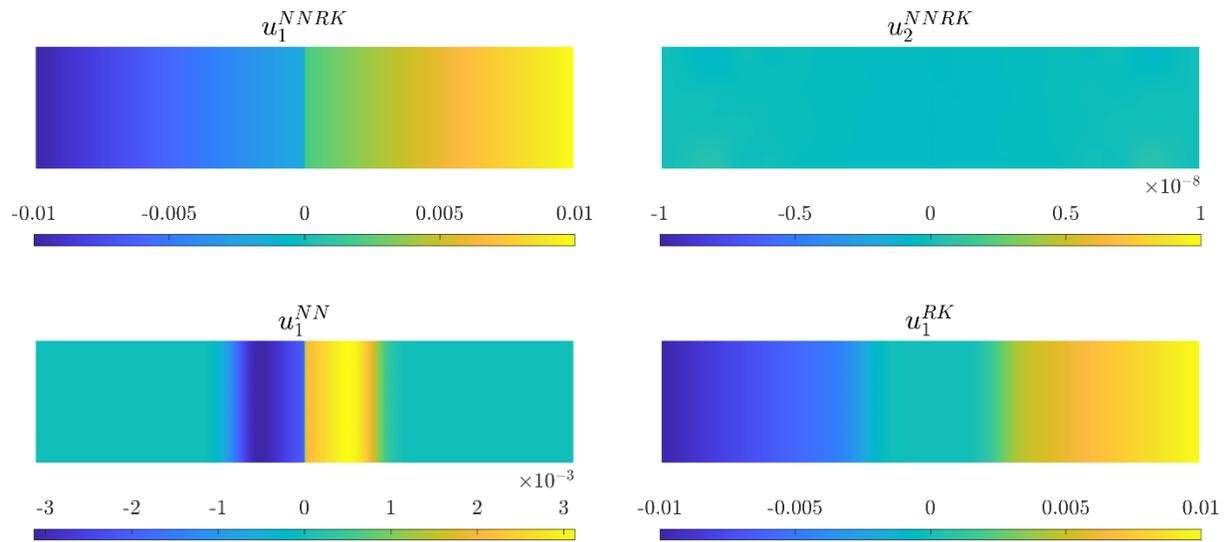

Figure 11. Predicted displacement (Case I)

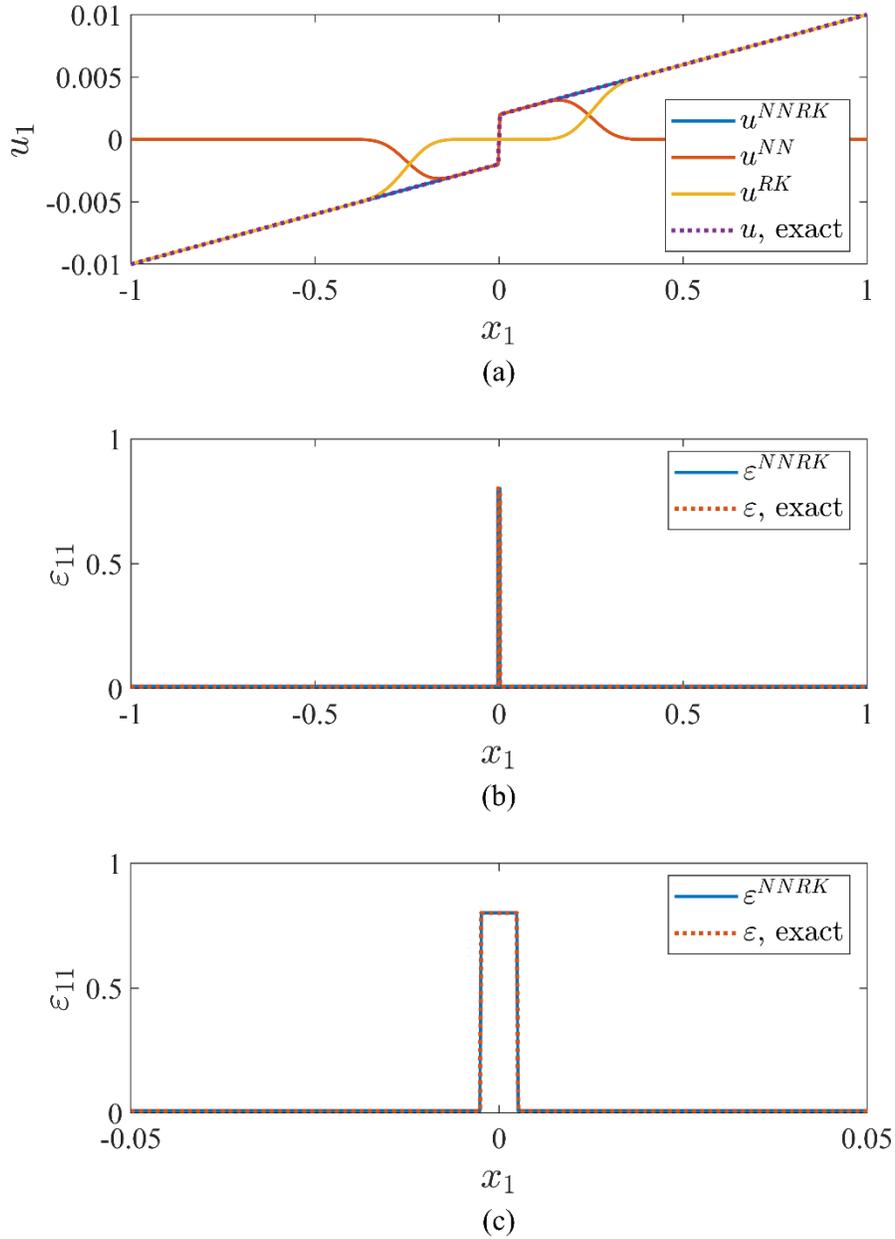

Figure 12. Numerical solution along $y = 0$ (Case I): (a) $u_1$, (b) $\varepsilon_{11}$, and (c) $\varepsilon_{11}$ (zoom-in)

For Case II, the background RK discretizations employed in this section are plotted in Figure 10 (a-c), and a 1,070,298-node, body-fitted Q8-FEM solution with a minimum nodal spacing of $H/2000$ near the localization (see Figure 13 for discretization) is used as a reference solution. Figure 14 shows the numerical solution for Case II, using $41 \times 11$ uniformly distributed background RK nodes (Figure 10 (b)) and a single 40-neuron hidden layer. Although the background RK discretizations shown in Figure 8 are relatively coarse compared to the width of degraded zone, the displacements predicted by the proposed method match the reference solution very well. The convergence curve for varying background RK nodal spacing ($h$) and the convergence curve for the varying number of neurons ($n_{NR}$) are plotted in Figure 15 (a) and (b), respectively. For the convergence study shown in Figure 15 (a), a fixed value of $n_{NR} = 160$ is used, and for the study shown in Figure 15(b), a fixed value of $h = H/40$ is used. Both results show convergence behaviors consistent with the error analysis result presented in Section 3.2.

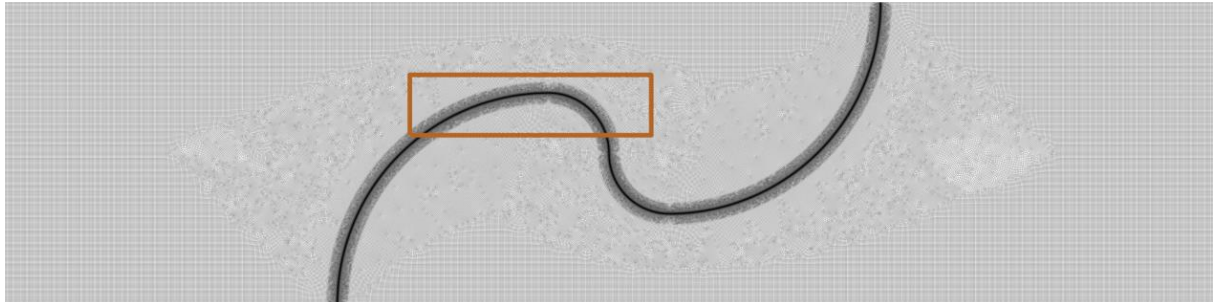

(a)

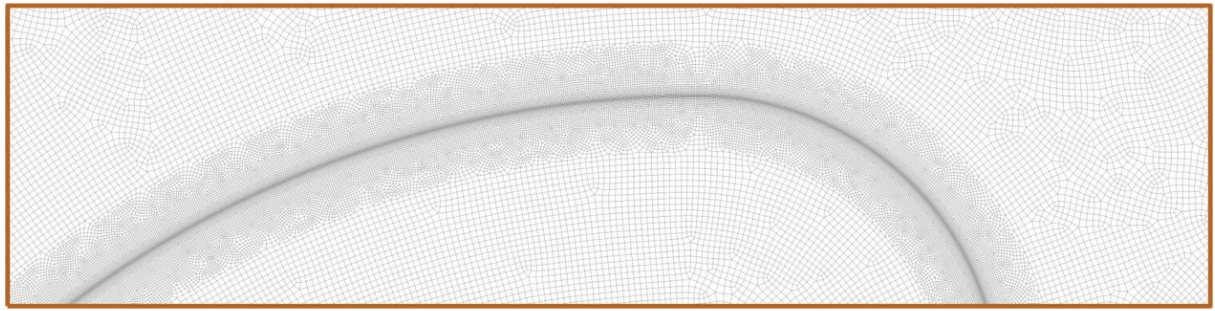

(b)

Figure 13. Body-fitted Q8-FEM discretization used to compute reference solution of Case II: (a) entire domain discretized by 1,070,298 finite elements with $h = w/12$ near the localization and (b) a zoom-in plot

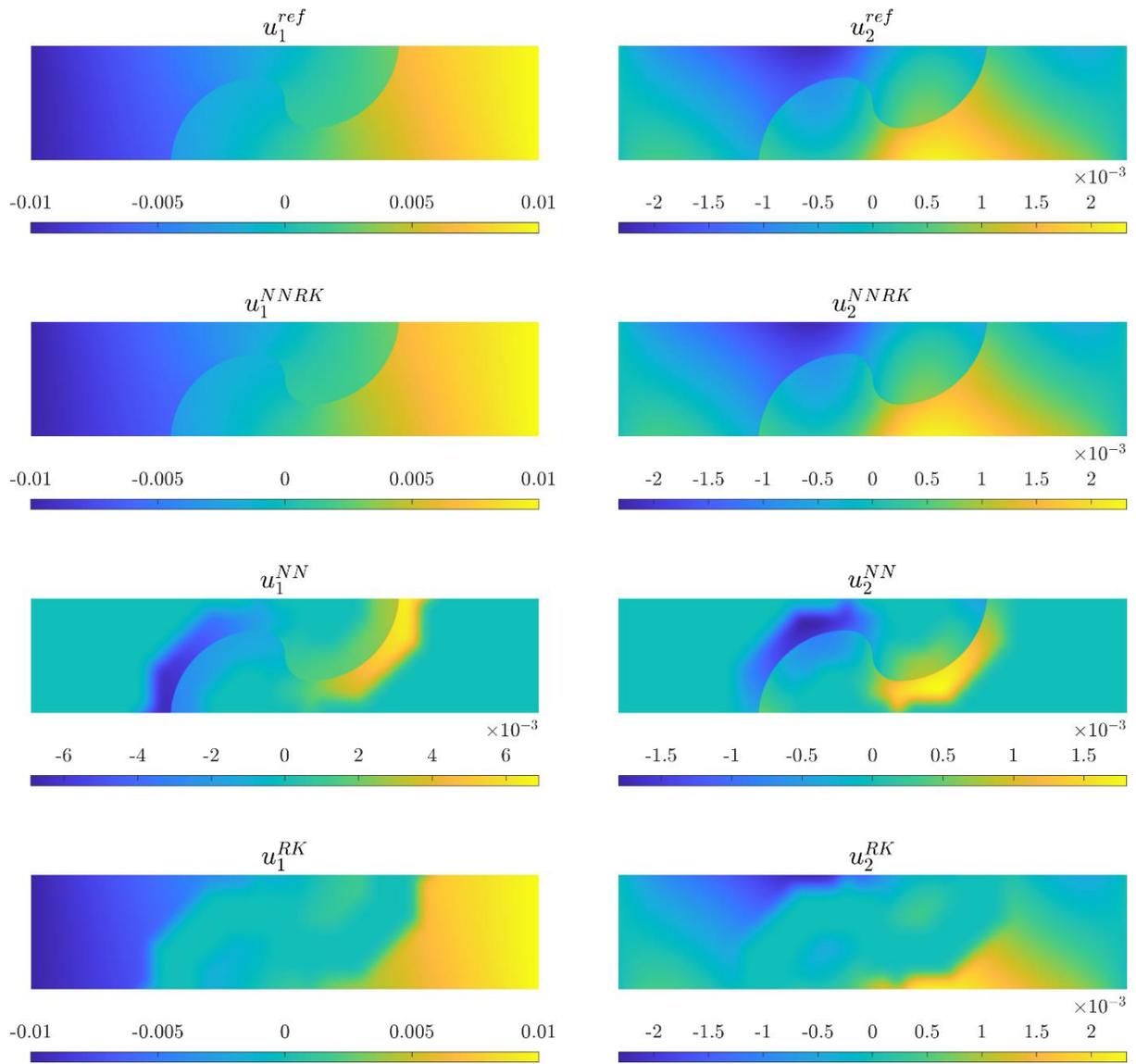

Figure 14. Displacement field (Case II): reference solution and NNRK solution (41×11)

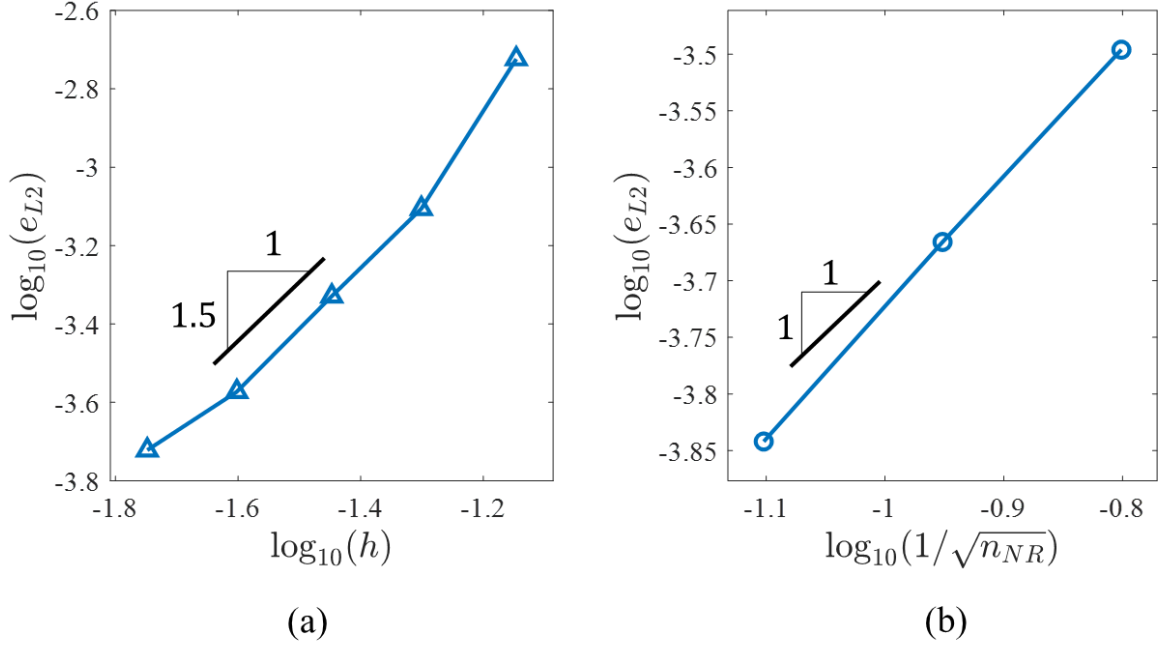

Figure 15. $L_2$ convergence rates: (a) for varying background RK nodal spacing with a fixed width of hidden layer ($n_{NR} = 160$) and (b) for varying $n_{NR}$ with a fixed RK discretization ($h = H/40$). The values enclosed by the parentheses in the legend denote the average convergence rates.

## 5.2. Pre-notched specimen subjected to simple shear

A benchmark problem of pre-notched specimen under simple shear is considered. As shown in Figure 16, a specimen with domain $\Omega = [-L, L] \times [-L, L]$ with a pre-existing crack of length $L$ is subjected to Dirichlet boundary conditions on the top and bottom surfaces. Specimen dimension $L = 0.5$ mm is used in this problem. The horizontal boundary value $g$ applied to the top surface is increased up to $15 \times 10^{-3}$ mm with an increment of $1 \times 10^{-4}$ mm. The material properties of $E = 210$ GPa, $\nu = 0.3$, $\mathcal{G}_c = 2.7$ N/mm are used. As shown in Figure 17, three levels of RK discretizations are used to study the regularization capability of the proposed method. For verification, a reference solution based on the reproducing kernel strain regularization [44] method is employed using 160,801 uniformly distributed RK nodes with nodal spacing of $h = L/200$.

Figure 18 (a-c) shows the damage propagation predicted by the proposed method. The damage is initiated with an orientation of approximately 65° and gradually changes the direction to the lower right corner during the propagation. The predicted damage paths plotted in Figure 18 (d) are not sensitive to the background RK discretization and agree very well with the reference solution. In addition, as shown in Figure 19, the load-displacement curves also demonstrate the good regularization capability of the proposed method and present reasonable agreement with the reference solution.

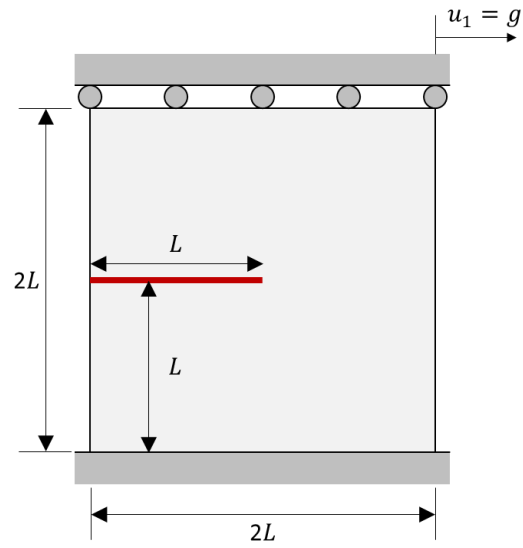

Figure 16. A pre-notched specimen for simple shear problem

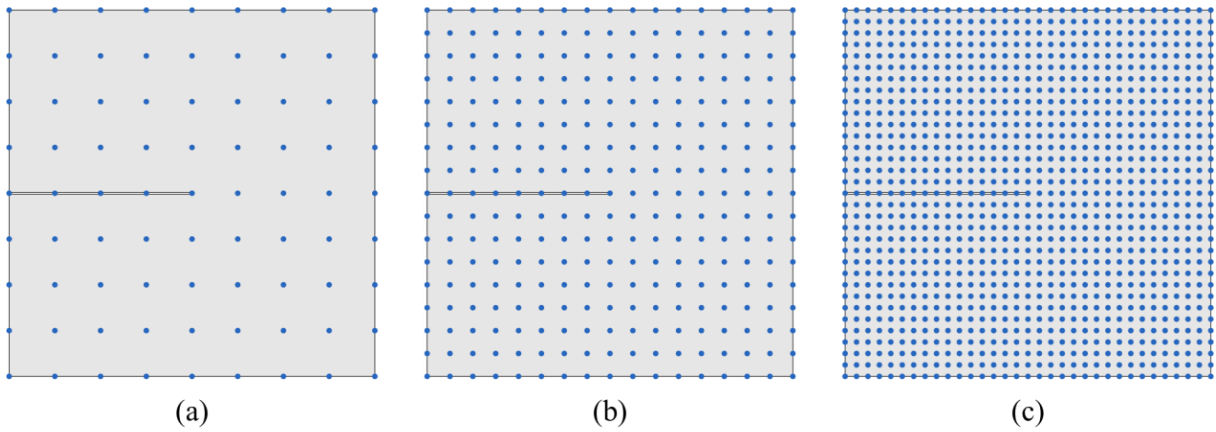

Figure 17. Background RK discretizations employed for simple shear problem. (a) M1: $h = L/4$, (b) M2: $h = L/8$, (c) M3: $h = L/16$

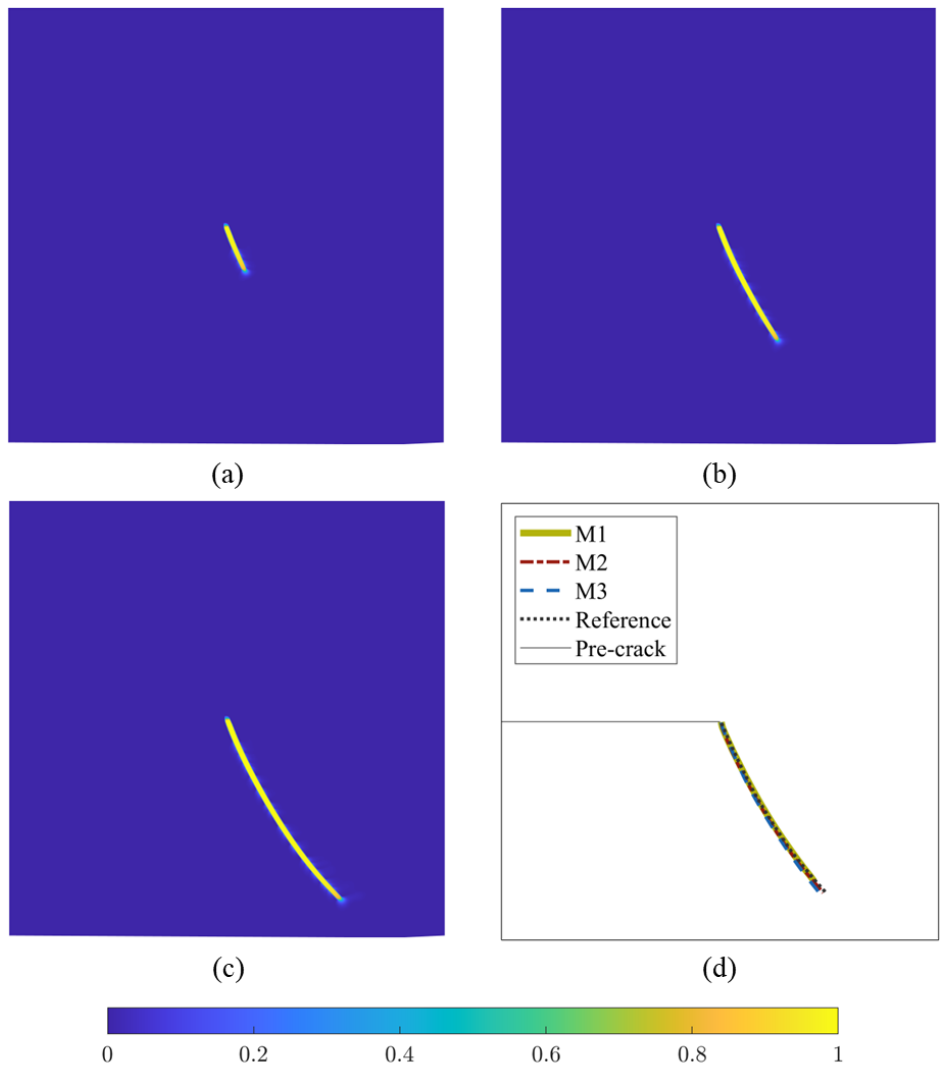

Figure 18. Damage evolution in simple shear problem (M2) for (a) $g = 9 \times 10^{-3}$, (b) $g = 10 \times 10^{-3}$, (c) $g = 11.5 \times 10^{-3}$, and (d) comparison of the predicted damage paths and the reference solution

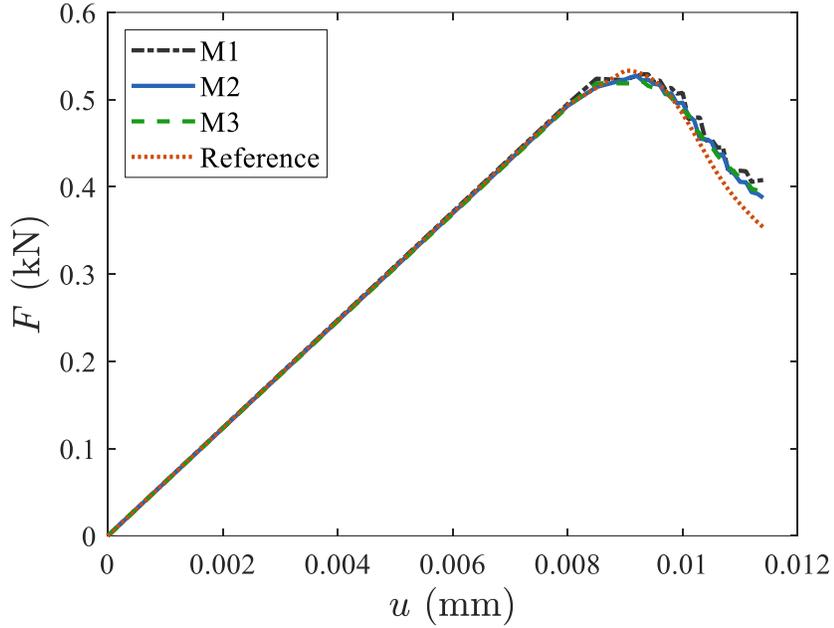

Figure 19. Load-displacement curve in simple shear problem

### 5.3. Quasi-static crack branching problem

In this section, the proposed method's ability to capture branching is demonstrated through a numerical example inspired by the problem proposed by Muixi et al [45,46]. Consider a square domain $\Omega = [-L, L] \times [-L, L]$ with a pre-existing notch with a length of $L$, as shown in Figure 20. The specimen is subjected to vertical displacement boundary conditions $g(x) = g_D(1 - x^2)/8$ on the top and bottom surfaces while the right surface is fixed in both directions. Herein, $L = 1$ mm is considered, and $g_D$ is applied up to 0.08 mm with $\Delta g_D = 4 \times 10^{-3}$ mm. The material properties $E = 20$ GPa, $\nu = 0.3$, and $\mathcal{G}_c = 8.9 \times 10^{-5}$ kN/mm are used.

In Figure 21, a progressive damage field is plotted in which the fracture initially propagates horizontally and branched near the fixed boundary as the accumulated strain energy associated with the vertical strain decreases due to the displacement constraint, which prevents further propagation of the fracture toward the fixed boundary. The branching is predicted to occur abruptly, then the propagation rate slows down. At the late stage of simulation, two branches switch the direction to the left. The overall trend of the damage propagation agrees with the reference PF-XFEM solution [46] superimposed in Figure 21 (d).

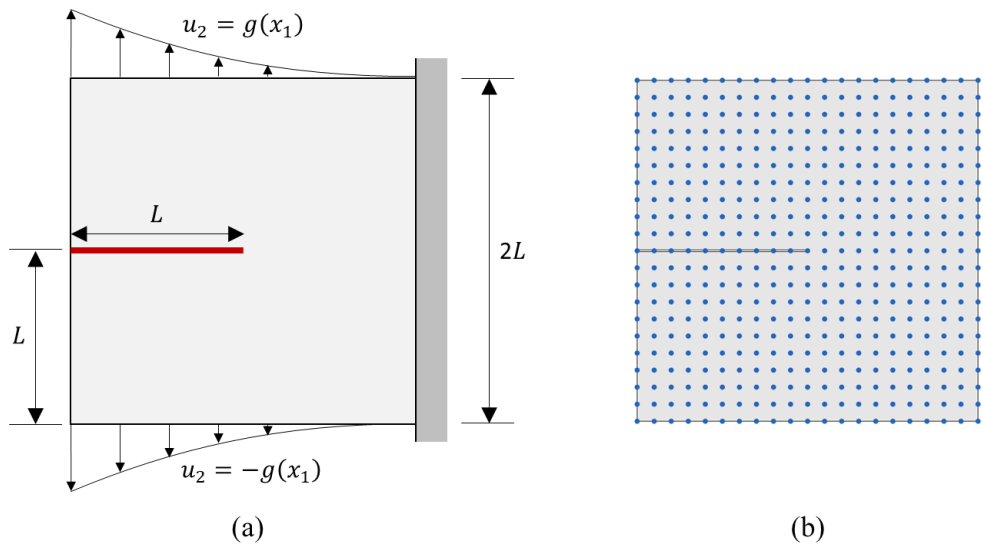

Figure 20. A pre-notched specimen for static branching problem: (a) geometry and boundary conditions and (b) background RK discretization

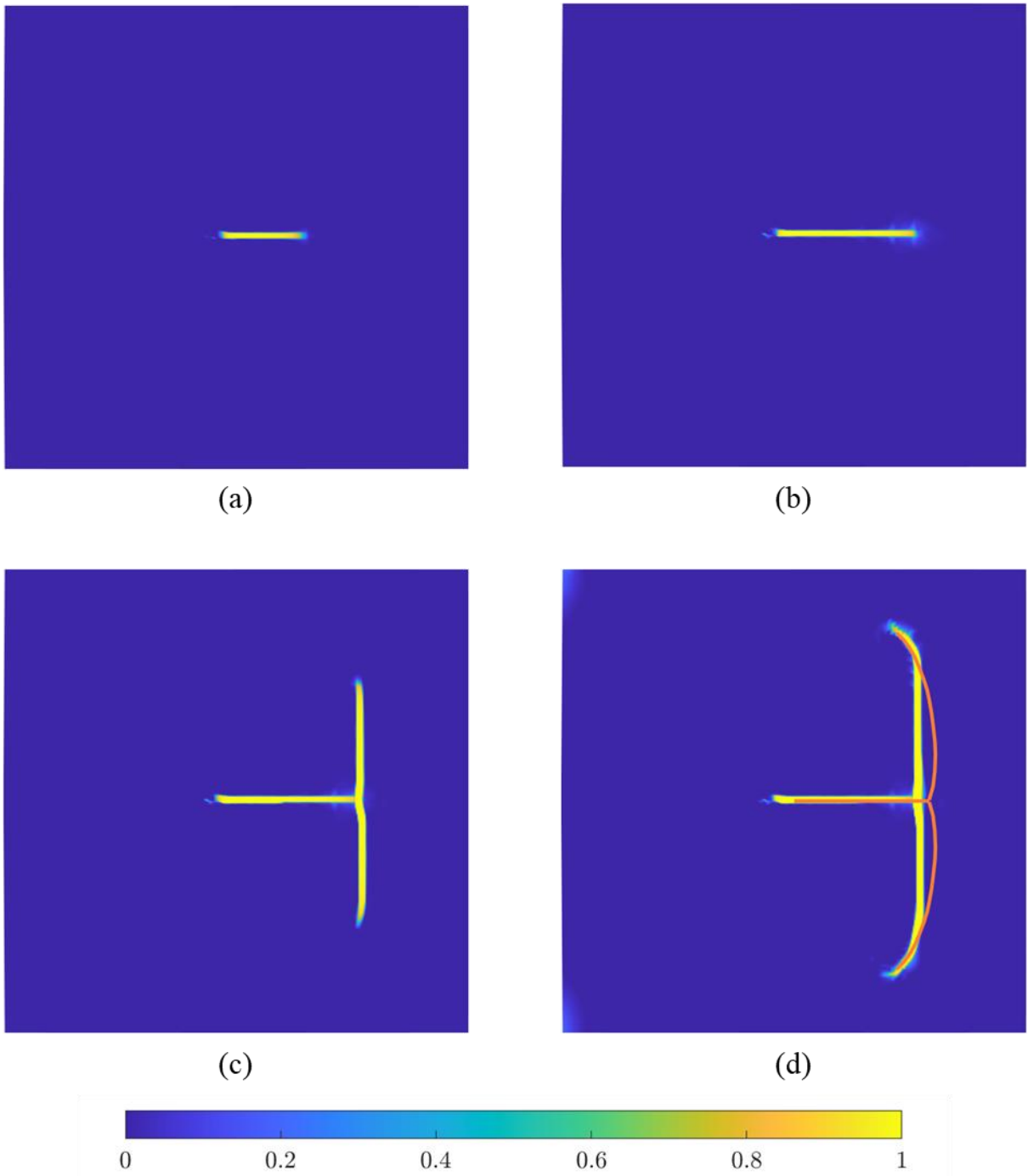

Figure 21. Predicted damage propagation and branching: $g_D$ of (a) 0.02 mm, (b) 0.036 mm, (c) 0.04 mm, and (d) 0.08 mm with a reference solution [46] superimposed in orange color

## 5.4. Mixed-mode fracture of a doubly notched rock-like specimen subjected to uniaxial compression

A uniaxial compression of a rock-like specimen with double pre-existing cracks [47] is simulated. As shown in Figure 22, a rectangular specimen with $H = 152.4$ mm consists of two 1-mm thick pre-existing cracks with $L = c = w = 12.7$ mm and $\alpha = 45°$. The Dirichlet boundary condition on the top surface is prescribed up to $g = -0.65$ mm with the increment $\Delta g = -1 \times 10^{-2}$ mm. Material parameters are Young's modulus of $E = 5.96$ GPa, Poisson's ratio of $\nu = 0.24$, the mode-I fracture energy of $\mathcal{G}_I = 5$ N/m, and the mode-II fracture energy of $\mathcal{G}_{II} = 20\mathcal{G}_I$. The domain is uniformly discretized by $16 \times 31$ RK particles. For NN approximation, the parametrization subblock consists of a neural network with two 40-neuron hidden layers along with the hyperbolic tangent activation function, which involves 1,842 unknown weights and biases. The NN length scale of 1 mm is employed.

Figure 23 shows the predicted damage propagation in the rock specimen. At the initial stage, four wing cracks are initiated from the four corners of the pre-existing notches and propagates with curved paths. Then, secondary shear cracks start to develop approximately at $g = -0.65$ mms the experimental observation [47].

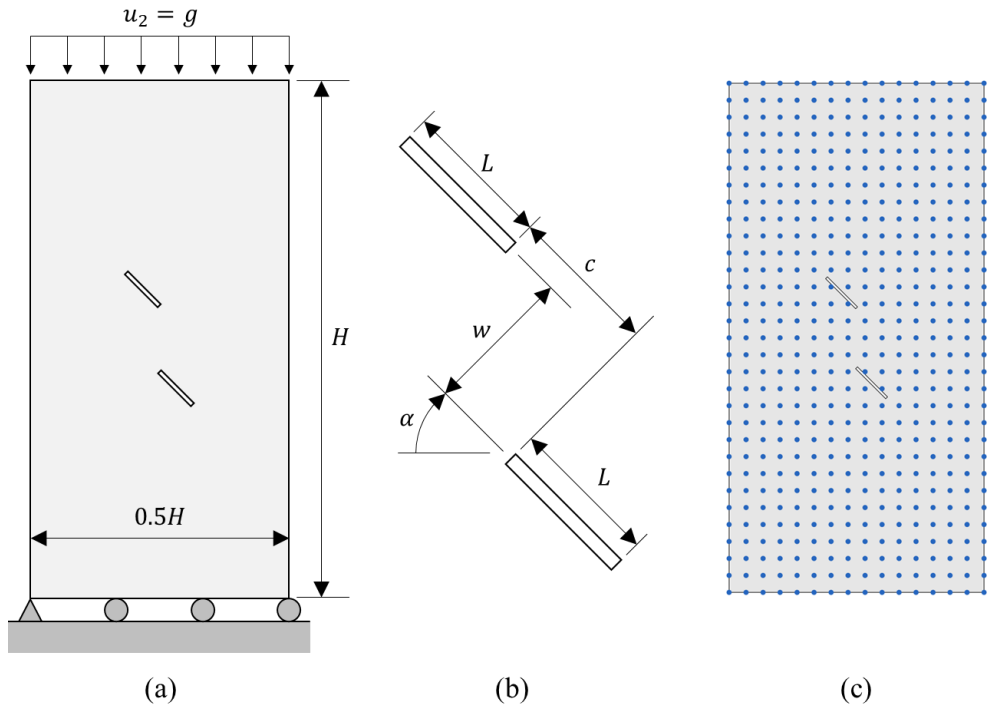

(a)  (b)  (c)

Figure 22. A rock specimen with double preexisting cracks: (a) geometry and boundary conditions, (b) details of preexisting notch, and (c) background RK discretization

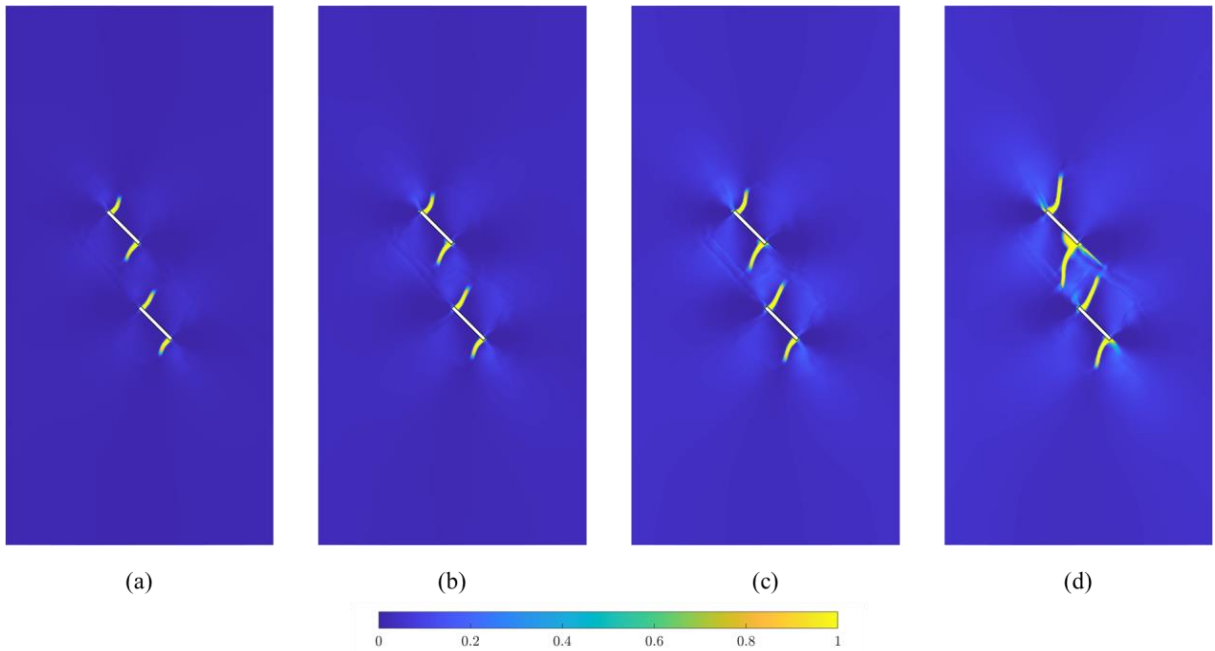

Figure 23. Progressive damage in rock-like specimen induced by uniaxial compression: $g =$ (a) -0.4 mm, (b) -0.5 mm, (c) -0.6 mm, and (d) -0.65 mm

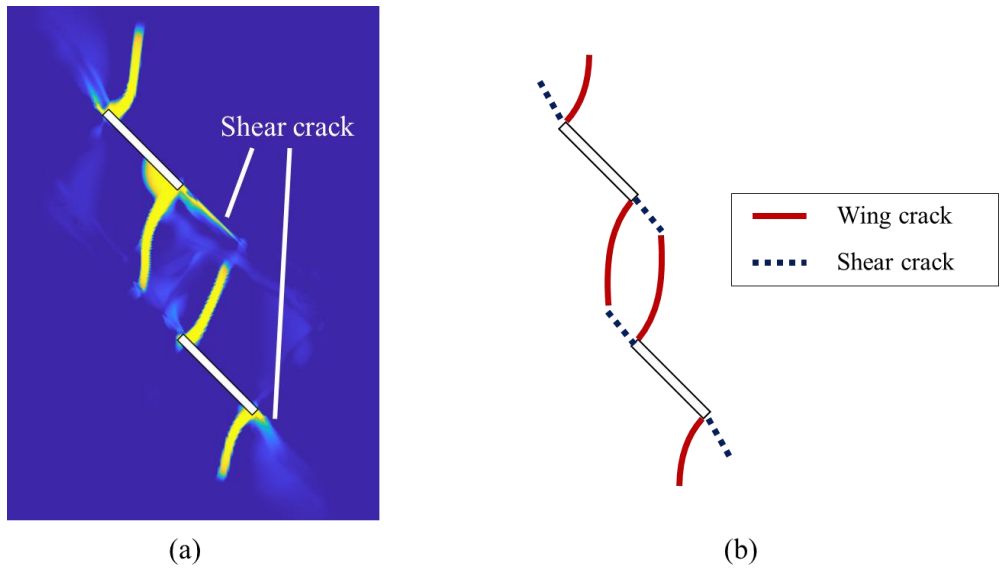

Figure 24. Comparison of (a) numerical results and (b) experimental observation [47]

# 6. Conclusion

An improved neural network-enhanced reproducing kernel particle method has been proposed for modeling brittle fracture. Derived through an NN-based correction of standard RK shape functions, the proposed method enriches a background reproducing kernel (RK) approximation with a coarse and uniform discretization by a neural network (NN) approximation equipped with a Partition of Unity property. The NN approximation is constructed by a deep neural network designed to capture localization, and the NN based enrichment functions are then patched together with RK approximation functions using RK as a Partition of Unity patching function. In the NN approximation, the deep neural network locates and inserts regularized discontinuities in the approximation function automatically, and the resulting NN enriched RK coefficient function provides varying magnitude of the discontinuities along the localization path.

To automatically capture the location, orientation, and solution transition across and along the localization, the optimum values of the control parameters contained in the deep neural network as well as the RK coefficients are obtained via minimization of the energy-based loss function. A regularization by introducing a constraint on the spatial gradient of the parametric coordinates to the loss function is employed to ensure a discretization-independent solution. Error analysis of the proposed NN-RK approximation is performed, and its verification with the numerical results show good agreement on the convergence rates. The numerical examples demonstrate the effectiveness of the proposed method in modeling damage evolution and branching with a fixed background discretization without conventional adaptive refinement.

# Acknowledgments

The support from the National Science Foundation under award #1826221 to University of California, San Diego, is greatly acknowledged.